\documentclass[preprint, 12pt]{elsarticle}
\usepackage[utf8]{inputenc}
\usepackage{graphicx}
\usepackage{subfigure}
\usepackage{color}
\usepackage{natbib}
\usepackage{textgreek}
\usepackage{upgreek}
\usepackage{multirow}
\usepackage{amsmath}
\usepackage{amssymb}
\usepackage{url}
\usepackage{hyperref}
\usepackage{comment}
\title{Initial Results From the First Field Expedition of UAPx to Study Unidentified Anomalous Phenomena}

\address[UA]{Department of Physics, University at Albany SUNY, Albany, NY USA}
\address[PX]{UAPx, Lake Panasoffkee, FL USA}

\author[UA]{Matthew Szydagis*}
\cortext[]{Corresponding author. \textit{E-mail address}: \\ mszydagis@albany.edu (Matthew Szydagis).}
\author[UA]{Kevin H.~Knuth}
\author[PX]{Benjamin Kugielsky}
\author[UA]{Cecilia Levy}

\begin{document}

\begin{abstract}

In July 2021, faculty from the UAlbany Department of Physics participated in a week-long field expedition with the organization UAPx to collect data on UAPs in Avalon, California, located on Catalina Island, and nearby. This paper reviews both the hardware and software techniques which this collaboration employed, and contains a frank discussion of the successes and failures, with a section about how to apply lessons learned to future expeditions. Both observable-light and infrared cameras were deployed, as well as sensors for other (non-EM) emissions. A pixel-subtraction method was augmented with other similarly simple methods to provide initial identification of objects in the sky and/or the sea crossing the cameras' fields of view. The first results will be presented based upon approximately one hour in total of triggered visible/night-vision-mode video and over 600 hours of untriggered (far) IR video recorded, as well as 55 hours of (background) radiation measurements. Following multiple explanatory resolutions of several ambiguities that were potentially anomalous at first, we focus on the primary remaining ambiguity captured at approximately 4am Pacific Time on Friday,~July 16: a dark spot in the visible/near-IR camera possibly coincident with ionizing radiation that has so far resisted prosaic explanation.~We~conclude~with~quantitative suggestions (3--5$\sigma$ rules) for serious researchers in the still-maligned field of hard-science-based UAP studies, with an ultimate goal of identifying UAPs without confirmation bias toward mundane / speculative conclusions.
\\
\\
\textbf{Keywords}: Unidentified Anomalous Phenomena, statistical significance

\end{abstract}
\maketitle

\section{Introduction: What is/are UAP(s)?}
\vspace{-4pt}
Recently, Unidentified Anomalous Phenomena \textit{i.e.}~UAP, formerly known as UFOs, have come under increasing scrutiny, with the U.S.~Defense Dept., U.S. Navy, and NASA taking them seriously. For decades, a culture of~mockery, pseudo-scientific claims, and media hyperbole created a stigma that prevented the mainstream scientific community from exploring this controversial topic. Now, in addition to our own collaboration, several academic research groups are all scientifically studying this subject~\cite{Teodorani,Villarroel_2021,Kayal+etal:2022,Kayal:2023,Watters_2023,Brothers}. In a collaboration with UAPx~\cite{UAPx},~UAlbany faculty have begun their own serious, technical study, part of a greater whole, to determine the nature(s) of UAP.

UAP can stand for Unidentified Anomalous Phenomenon or Phenomena, although the A can also stand for Aerial/Aerospace; UAP was recently redefined by the U.S.~Congress as Unidentified Aerospace-Undersea Phenomena. The term aerospace broadens the study of UAP to include the Earth's atmosphere and outer space; whereas the term undersea extends it to underwater and oceanic domains. The importance of the multi-medium nature of UAPs is perhaps best illustrated by the fact the successor to the UAP Task Force is the All-domain Anomaly Resolution Office or AARO, which was established by the U.S.~Defense secretary to provide ongoing reports to Congress. UAP have come to their attention as an old, global air-safety risk at least~\cite{ESCOLAGASCON2024102617,NewZealand}.

The term UAP refers to an object/phenomenon which cannot be immediately recognized as prosaic, \textit{e.g.}~for aerial~phenomena: human-made craft or flying animals. Unidentified only means that one does not know~what~something is -- at least initially -- until additional analysis is possible. While~many UAP are ultimately identifiable, some remain unidentified. It is important~to note that UAP is a broad classification. It is unlikely all UAP, especially those resisting classification, stem from a singular phenomenon, and identification is not binary. It depends on how much desired detail is possible by means of a given sensing method and the descriptive categories brought to bear on it.

There is little doubt that the majority of UAP are misidentifications,~but anywhere between 4-40\% remain unidentified after careful investigations~\cite{Clemence:1969,SturrockVCondon,Cometa:1999,AirForce:2003,Davidson:1966}, depending upon the sources and quality of the reports. Despite this, there exist hard data that demonstrate unreasonably high speeds (above Mach~40-60) and accelerations (thousands of times \textit{g}) \cite{Oberth:1954, Hill:1995, Poher:2005, Hill:2014, Knuth+etal:2019, Coumbe:2022}, without corresponding sonic booms or fireballs. Such observations represent cases of interest as they support the more exotic hypotheses, requiring novel physics or at least new engineering; but, more data are needed to characterize fast-moving objects and definitively rule out observational errors~\cite{SonicBoom}.

While making an exhaustive list isn't feasible, the possible explanations~of UAP include, but are not limited to: airplanes, helicopters, drones,~balloons, satellites / ISS and other human-made aircraft (or, spacecraft) such as para-gliders / para-sailors, and marine vessels. Fauna were already mentioned, but natural explanations can also include -- clouds (especially lenticular),~atmospheric optical effects, such as sun dogs, solar / lunar halos, Fata Morgana optical illusions and other types of mirages, plus celestial bodies, like the Moon, Venus or other planets, meteors, and comets. Many different~phenomena are likely being included under the UAP umbrella. Heretofore non-discovered or sparsely studied atmospheric phenomena may be involved \textit{e.g.}~ball lightning \cite{Cen+etal:2014, Singer:2012} or earthquake lights \cite{Freund:2003,St-Laurent+etal:2006,Freund:2019}. Both of those have been considered ``pseudoscience,'' despite well-documented observations to the contrary~\cite{Stenhoff:1999}.

It is this vast number of prosaic explanations, both known and unknown, which makes identification a major challenge. However, enough reports~from serious, legitimate sources (\textit{e.g.}~\cite{Cometa:1999}) clearly highlight a need to study UAP scientifically, while reducing biases. Groups such as UAPx \cite{UAPx}, IFEX~\cite{Kayal+etal:2022,Kayal:2023,GermanAnoNewest}, VASCO~\cite{Villarroel_2021}, and Galileo Project~\cite{loeb2022overview} are each undertaking this task now. Villarroel and her group (VASCO) have already made seminal discoveries of concrete anomalies, multiple light sources appearing and vanishing~\cite{Villarroel_2021,solano2023bright}.

\subsection{What is UAPx?}

UAPx is a non-profit organization, co-founded by veterans Gary Voorhis and Kevin Day, who were involved in the (2004) \textit{Nimitz} carrier strike group UAP encounters~\cite{Powell+etal:2019}. While not first to collect field data it is one of the~first organizations of its kind to do so in the era after the watershed moment~of the clip released of the \textit{Nimitz} / \textit{Princeton} ``tic-tac'' and other videos. UAPx was founded at a similar time as Galileo Project, with similar aims. UAPx recruited scientists, engineers, and technicians to work alongside experienced veterans for the purpose of collecting and scientifically analyzing new UAP data. This is distinct from re-analysis of historical UAP cases. That work~is already done by multiple individuals~and organizations: $e.g.$ SCU (Scientific Coalition for UAP studies~\cite{SCU}), plus MUFON (Mutual UFO Network~\cite{MUFON}). Some cases have included hard evidence, summarized (for those up to 1998) authoritatively by \cite{Sturrock:2000}, which cited and built upon \cite{Rodeghier:1981} and \cite{CashLandrum}, and others. Existing groups diverge significantly in amount of scientist participation and rigor. The approach taken by UAPx ensures data quality, chain of custody, data provenance, and rigorous analysis, but having independent groups take that approach is useful for cross-checking claims, as seen in other fields.

Scientists collaborating with UAPx constitute a wide range of differing expertise and informed opinions; they strive for an introspective self-skepticism, aiming for the difficult balance between openness to speculative ideas~on the one hand~\cite{Vallee}, versus a debunking posture on the other, aimed exclusively at explaining all ambiguities as part of known phenomena regardless of context.
\vspace{-17pt}
\subsection{The Objectives of UAPx, and Its Motivations}

UAPx is devoted to identification and classification of the initially unidentified and unclassified. While open to the techno-signatures possibility, as astronomers should be~\cite{villarroel2023astronomical}, including near-Earth signatures, it neither excludes mundane options nor prefers speculative ones, respecting Sagan's dictum regarding extraordinary claims and evidence, but not blindly. Its goals are:

\vspace{-5pt}
\begin{itemize}
    \item Acquire data which fill the gap in the scientific knowledge of UAP due to a lack of high-quality, unclassified data.
    \vspace{-10pt}
    \item Make said data / analyses of them publicly available. Analytical results and raw data will be released only after analyses are completed. Completion is defined as peer-reviewed publications, in scientific journals.
    \vspace{-25pt}
    \item Combine existing hardware in new ways as well as develop new sensors as needed, in order to accomplish the first two goals.
\end{itemize}
\vspace{-5pt}

In service of all three objectives, UAPx has developed, and continues~to~refine, methods for collection of quality data on UAP, through use of a small-scale (portable, tabletop, lightweight) but diverse (as advocated~by~\cite{Beck}) and advanced, multi-spectral suite of sensors, covering not only the EM spectrum, but also E/B-fields~\cite{Meessen2012EvidenceOV}, ionizing radiation, and other types of measurements, made robust by calibrations before/during/after field deployments, per \cite{Teodorani}.

From July 10-17 of 2021, UAPx conducted an expedition using two locations, in Laguna Beach, CA on a flat rooftop, and in Avalon, CA on Catalina Island. (The timing reduced the probability of a fully cloudy week.) A third, moving viewpoint was a 1999 Land Rover, the O.S.I.R.I.S.~or Off-road Scientific Investigation \& Response Informatics System~\cite{OSIRIS}. As will be~delineated within the instrumentation section (3), the team deployed 1 UFODAP (UFO Data Acquisition Project) dual-camera system with 1 PTZ (Pan Tilt Zoom) and 1 fisheye camera (although this fisheye was separate and always on the O.S.I.R.I.S.~roof), 2 pairs of night vision (NV) goggles for use by the island team, 8 FLIR (Forward-Looking InfraRed) cameras, and 1 Cosmic Watch, a radiation detector. (Other tools were less useful, and thus they are not listed here; furthermore, note that the full setup was not complete until July 11.)

The 2004 \textit{Nimitz} case~\cite{Knuth+etal:2019} triggered the collaboration's interest, although the Catalina region was a purported hotspot before that occurrence~\cite{TCI:2013,PDennett}, and before it became public knowledge in 2017 (The \textit{New York Times}).~The location was essentially pre-selected, due to funding being provided for travel to only that one, but rationale for the locales of later expeditions should come from studies like \cite{MedinaSciRepKirk}, which was not published when UAPx was formed and planning its first field expedition, assuming that population density and similar factors are taken into account to identify hotspots. (See also~\cite{France}, similar to \cite{MedinaSciRepKirk}, for France, with references to its agencies GEIPAN and CNES.)

A reason for two (or more) observation locations in the same general~area is to have multiple, unique vantage points for the same UAP in an encounter. Another is to avoid past criticisms of one poor-quality image in one camera. If sites are close to one another but sufficiently separated,~\textit{i.e.}, far beyond the position resolution of the measuring devices (but still near the hotspot, unlike \cite{Twinkle}), triangulation is feasible, as attempted by Maccabee~\cite{Maccabee}. While this was vigorously pursued, the first expedition did not succeed in such data fusion. Due to the expense of transport of bulky materials, it was not practical for islanders (communicating via phones only) to have anything more than NV (two pairs). We suggest having only practical, identical setups going~forward.
\vspace{-21pt}
\section{Methodology and Context}
\vspace{-2pt}
Watters~\textit{et~al.}~\cite{Watters_2023} describe the challenges faced for UAP studies in great detail, explaining how and why we can build on ``gray'' literature (non-science publications) including anecdotal evidence, for sensor choices.~They~establish global objectives for this kind of science, and provide a framework and~a~road map for discoveries. In addition to providing a foundation to build upon, \cite{Watters_2023} already addresses a need for controls and background data, bias mitigation, and reproducibility, and contains a lengthy bibliography of preceding~instrumented studies. Combined with the moving away from eyewitness testimony, these tools should help move the study of UAP out of the ``fringe.''

UAPx uses a diverse set of devices to capture different types of data on many channels. The strategy of the UAPx / UAlbany collaboration involves:

\begin{enumerate}
\vspace{-4pt}
\item Acquiring and commissioning multiple cameras and multiple copies of other sensors, to capture the same phenomenon from different angles, at high resolution, for robust estimations of distance, size, speed, and acceleration by triangulation, possible via precise unit~locations~\cite{Szenher_2023}.
\item Coincidence timing across all devices to help in a faster data reduction resulting in lists of ambiguities and true ``anomalies.''
\item Eventual construction of (semi-)permanent sites of automated remote sensor suites, to serve as (near-continuous) data collection~stations (like \cite{MADAR}), in control areas (in parallel, for a large enough group) and supposed~UAP hotspots, inspired by \cite{Hessdalen2001,Hessdalen2013}. A similar approach~to~anomalous atmospheric luminous phenomena was already considered in \cite{TeodoraniAilleris}.
\item Corroboration of anomalies via use of public sources of data, including but not limited to Doppler weather radar data, any unclassified satellite imagery, as well as global particle/radiation detection networks.
\end{enumerate}

To serve points (2) and (4), Monte Carlo simulations and both Bayesian and frequentist inference will be applied to determine the probability of two or more events overlapping in time being due to accidental coincidence. AI/ML (Artificial Intelligence / Machine Learning) will also play an instrumental role in UAPx, for a non-binary and multi-stage classification, beginning with motion detection, for infrared (IR) and for visible light. In the first expedition (Section~4), (1) was only partially successful, as only one non-IR camera was permanently running, (2) was hampered by imprecise timing, to be discussed later as a lesson learned, (3) is a longer-term aim, and (4) is underway.

We must place UAPx and its mission in its context within the history of UAP / UFO studies, comparing and contrasting it with both contemporary and historical endeavors. We aim for a presentation of several representative examples of teams focused upon fresh, non-human-eye data collections. For~a more comprehensive review (of efforts since 1950) one is encouraged to read Ailleris' article~\cite{ailleris2024exploring} in \textit{Limina: Journal of UAP Studies} which includes UAPx in its list, in addition to Stahlman~\cite{GStahlman}.

* \underline{The Galileo Project (GP)}

At time of writing, the GP and its affiliated researchers can lay claim to having the most UAP-related papers published the most recently in a high-impact, peer-reviewed journal (\textit{JAI})~\cite{Cloete,loeb2022overview,loebOcean,Mead_2023,Randall_2023,Szenher_2023,Watters_2023}. But,~GP~only has presentation of instrumentation and analysis as plans. It has no data~yet from an expedition out in the field, except from the search for an interstellar meteorite on the floor of the Pacific Ocean~\cite{loebOcean}. Such searches, as well as ones for USOs (Unidentified Submersible Objects)~\cite{PDennett,ITSanderson} are beyond the current scope, abilities, and budgetary profiles of UAlbany / UAPx. Moreover, another difference is GP is yet to test instrumentation at alleged~hotspots.

GP and UAPx do not differ greatly in terms of detectors~\cite{Watters_2023}, but novelty is less important than independent replication,
since reproducibility is the hallmark of a successful experiment.

* \underline{VASCO and EXOPROBE, plus ECRI}

VASCO (Vanishing and~Appearing Sources during a Century of Observations) has already made detections~\cite{villarroel2023astronomical,Villarroel_2021,villarroel2022background}. It is complementary to UAPx, seeking transitory lights in pre-Sputnik photographic plates from observatories. A lesson we can take away from VASCO~is~citizen scientist engagement for review of images. EXOPROBE~\cite{ExoProbe} will continue the same style of searching for non-human probes via a global optical-telescope network, while ECRI (the European Crash Retrieval Initiative)~\cite{ECRI} also presupposes physical craft, ones capable of crashing~\cite{NOLAN2022100788}.

* \underline{IFEX}

IFEX -- Interdisciplinary Research Center for Extraterrestrial Studies~\cite{IFEX} is based in Germany. Like GP, it deploys sensors like rooftop cameras~(Sky-CAM-5) first locally, but also launches its own satellites (SONATE) and has potential to see UAPs from above~\cite{Kayal:2023}. IFEX has boldly merged UAP studies with SETI. While some members of UAPx are also affiliated with IFEX, we remain more agnostic regarding any association between UAPs and possible ET(I), giving comparable credence to exotic natural phenomena as potential explanations, yet still acknowledging more exotic possibilities~\cite{Magnet}.~It~is~also important to note that -- just like VASCO and ECRI -- the work of IFEX~underscores the international nature of the UAP question.

* \underline{The Nightcrawler}

Nightcrawler is a large vehicle for mobility, with a slew of onboard sensors \cite{Brothers}. Like UAPx, its creators have thus far only gone to one potential hotspot (Long Island, NY). The sensor suite (visual, IR, high-energy particles) is very similar to UAPx's, but adds passive radar, one serious deficiency of ours, due to cost and an initial lack of expertise.

* \underline{Project Hessdalen}

The Hessdalen valley in rural central Norway has becomes famous for an inexplicable, bright phenomenon, studied for years with no conclusive~statements made regarding its true nature~\cite{Teodorani,Hessdalen2001,Hessdalen2013}. Two lessons we glean from this project are choice of a superior locale, known for reproducible UAP, and spectroscopy for remotely determining elements and distances, done for the Marfa, TX lights too~\cite{Marfa}.

* \underline{Project Identification}

Published in a book like the Tedescos' work~\cite{Brothers} (at first), that of physicist Prof.~Rutledge, initially skeptical and critical, claims to be the first scientific field study of UFOs~\cite{Rutledge}. Modern attempts owe much to him, and one cannot do his study justice in the space which we have here. Together with students, he observed dozens of sightings not easily explainable by \textit{e.g.}~planes by means of a camera, telescope, and binoculars, suggesting that ``The Phenomenon'' is something real and worthy of study. They were hampered by unclear (B\&W) photography and no automation, not yet prevalent at the time.
\vspace{-7pt}
\section{Instrumentation / Equipment List}
\subsection{Visible Light Imagery (UFODAP)}

For visible / near-IR imagery, UAPx selected, for its first expedition, the UFODAP~\cite{UFODAP}, chosen due to its advertised ease of use, data security, and an optional collection of secondary sensors. The PTZ camera, capable~of~physically tracking potential objects of interest through the UFODAP's OTDAU (Optical Tracking Data Acquisition Unit) software, utilized a commercial~machine vision algorithm to identify moving targets and track them by rotation of the lens as needed. It was left at its default high sensitivity for the 07/2021 Catalina mission, to avoid missing any potential anomalies.~(One~example of a UFODAP alternative is Sky360 \cite{Sky360}, formerly called SkyHub; the UFODAP project is partnered with UFODATA~\cite{UFODATA}.)

The UFODAP does not record without a trigger, so it does not overwhelm storage capacity with non-stop, high-quality video. However, this comes with a downside: if the trigger isn't well set up, only a few frames may be captured for ambiguous events and/or reams of frames may be triggered by mundane motion. That makes frames difficult to analyze, especially for objects~that~are unexpected (classic UAP) and difficult for the OTDAU to detect.

Between 22:35 Pacific Sunday July 11 and 12:40 on Friday July 16, 2021, 1716 AVI files were saved from the UFODAP, with the majority $O$(1) second in length, or even shorter (a few frames in many cases). Because of metadata blocking portions of frames, two versions of each of the 1716 files exist, both with / without reticles, date, temperature, angle, and other information, for the purposes of having both unobstructed views plus all relevant metadata.

\begin{figure}[ht]
\centering
\includegraphics[width=0.99\textwidth,clip]{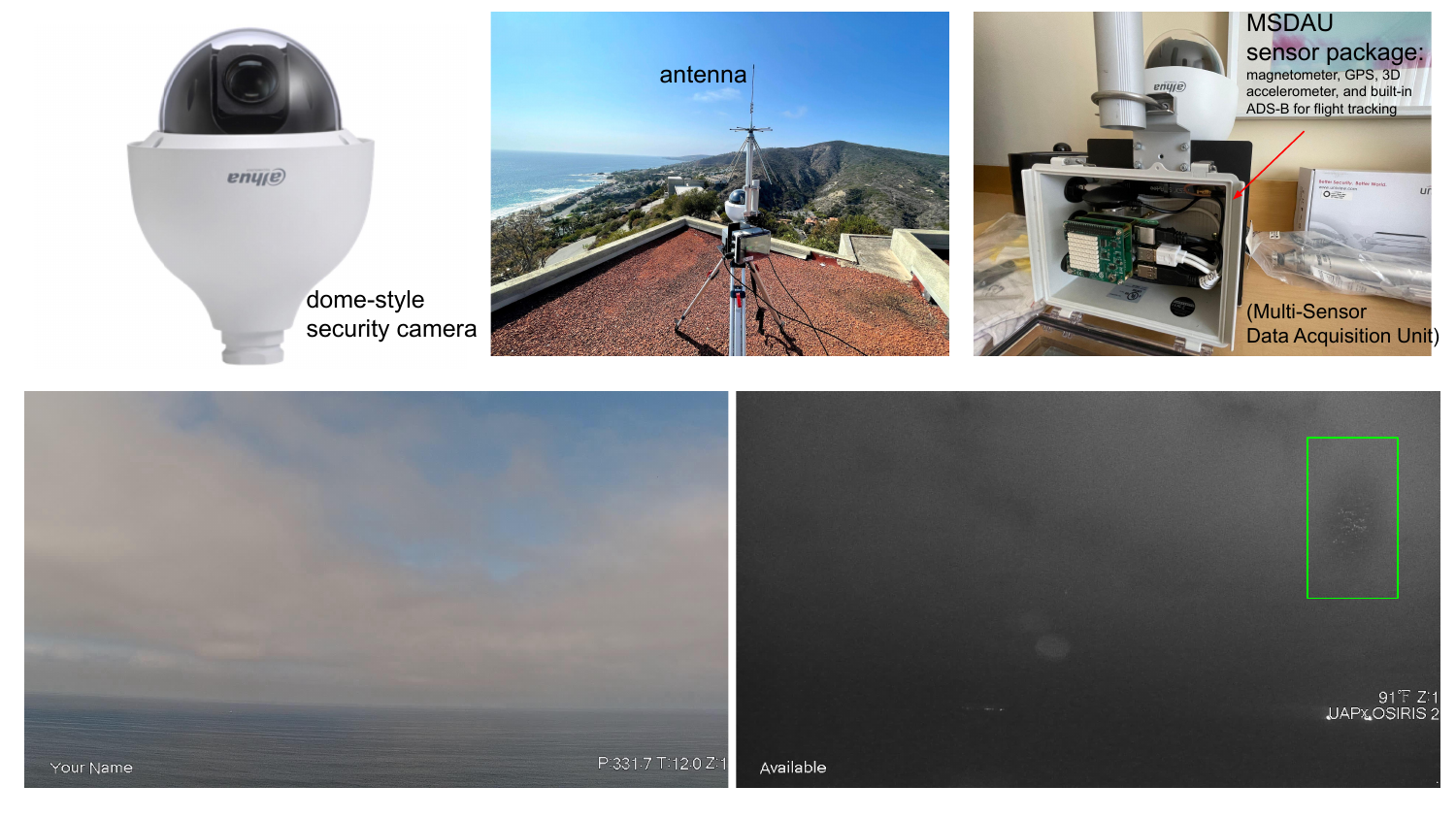}
\caption{Upper left: Example Dahua 50232 camera at the heart of the default UFODAP system. Upper middle: A UFODAP unit deployed on a roof in Laguna Beach, California in mid-July, 2021. (The discone antenna is for RF spectrum analysis, but it was uncalibrated and has not yet been used by UAPx.) Upper right: A sensor package that came with the UFODAP for UAPx, with the cover removed to display the internals. Its software was not functional during the first field expedition, but sensor data will be recorded in the future. Bottom left: An example daytime UFODAP image of the Catalina channel, at 1920x1080 resolution. Bottom right: The view from the same perspective but at 4am. In this sample, a dark spot of unknown origin with bright pixel clusters within is present (green box).}
\vspace{-0pt}
\label{Fig1}
\end{figure}

The top row of Figure~\ref{Fig1} displays the physical system from different angles and in different contexts. The bottom row presents two sample images, of the same field of view, at the default orientation (prior to any PTZ rotations), at different times of day. The observation point was within Laguna Beach, CA with the field of view centered on an azimuth of 260.5$^{\circ}$, providing a view of Catalina Island across the channel. The lower right has a single video frame from the most perplexing ambiguity captured during UAPx's first expedition.

\subsection{IR: Night Vision and FLIR}

The utility of infrared videography has been demonstrated by the \textit{Nimitz} \cite{Powell+etal:2019,Knuth+etal:2019}, \textit{Roosevelt}, and other U.S.~military incidents. Not only does IR video-graphy provide an ability to capture images at night, but it also provides data on the temperatures within the field of view, and can reveal the presence, or absence, of high-temperature exhaust. That said, for all of the IR equipment we only possessed the existing internal calibrations, and thus our devices were more useful for measurements of relative not absolute temperatures (simply cold vs.~hot), and for contrast against the background.

\begin{figure}[ht]
\centering
\includegraphics[width=0.95\textwidth,clip]{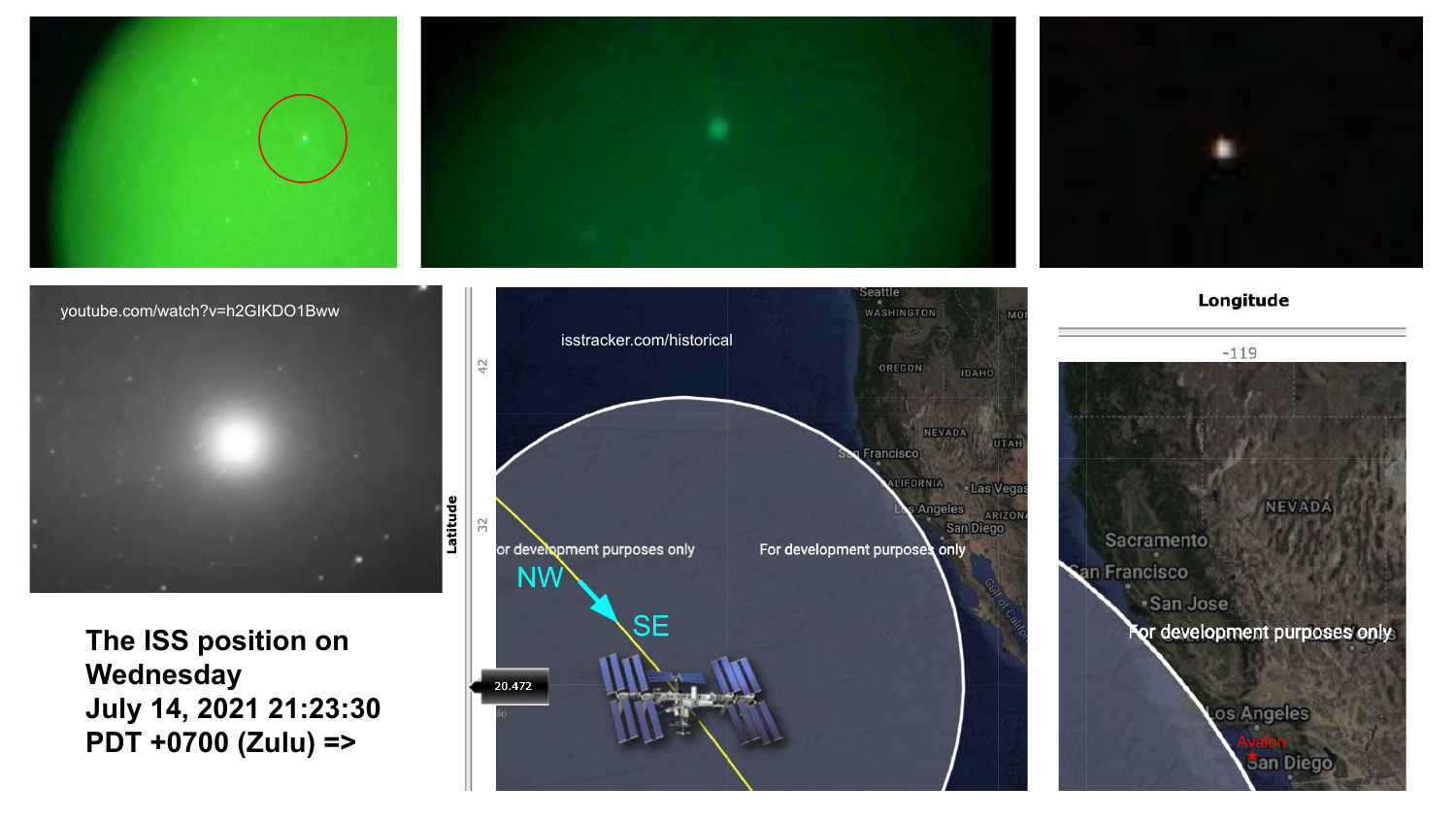}
\vspace{-7pt}
\caption{Upper left: A video frame (720$\times$1280) showing a spheroid in night vision (phone recording). It neither blinked nor made noise, but was later identified as most likely being the International Space Station or ISS (see Appendix 1 for technical details). Upper middle: An example from another phone (480$\times$640) that captured the same object, concurrently. Upper right: A movie camera belonging to the documentary crew also recorded it. Bottom left: Online example of the ISS in night vision, for a qualitative comparison. Bottom right: Details on the ISS location at the time (site used in beta-test mode). The white curve denotes the edge of a nominal 80-km-radius horizon, but Avalon on the island (where the videos were shot) could still see the ISS for $\sim$120 s due to altitude and weather.}
\vspace{-9pt}
\label{Fig2}
\end{figure}

UAPx employed several pairs of AN/PVS-7x night-vision goggles, usable for manual spotting, as one could not record (directly). Smartphone cameras held up to the night-vision goggles did provide an elementary recording capability for potentially interesting events (example in Figure~\ref{Fig2}, now explained).

The primary imaging systems used on the expedition were 8 FLIR ThermaCam PM695 units. These were deployed on the mainland (Laguna Beach) on the same flat roof as the UFODAP unit discussed in the previous section, and pointed in the direction of the Catalina channel. They were oriented~with slightly overlapping fields of view, to allow for triangulation -- although with large uncertainties due to the units' proximity to one another. They recorded imagery using a 720$\times$360 uncooled microbolometer array with spectral sensitivity from~7.5-13~$\mu$m and field of view of 48$^\circ$$\times$36$^\circ$. The units were operated at 60 fps~and the resolution was up-scaled in software to save files as 1440$\times$720 MP4s with two 4-channel AJA Ki-Pro-Go digital video recorders. Via FFmpeg~\cite{FFMPEG} as the first step, videos were analyzed frame by frame, at the higher resolution, with human-eye quality assurance on motion detection of discrete pixel clusters moving across the screen combined with threshold variations. The image frames were cropped to address (bouncing) cursor misbehaviors. (The FLIR cameras were deployed near Seattle before this study for years.)

\subsubsection{C-TAP FLIR Image Analysis}
\vspace{-1pt}
A new software package was developed by Szydagis for rapid image processing of FLIR recordings which has been dubbed C-TAP (Custom Target Analysis Protocol), capable of detecting discrete objects traversing a screen, even in conditions of rolling clouds, by means of a na\"ive Bayesian classifier applied primarily to the minimum and maximum differences in pixels between frames, and the standard deviation across RGB pixel values. This helped differentiate (physical) objects captured crossing the fields of view from camera noise. Only color-mapped RGB values were available, not mono-channel intensity data from the microbolometers within the PM695s. This suggests the ability to distinguish noise using the RGB standard deviations was an~effect from the internal software a PM695 uses to render temperature colors. Such an effect may be unique to that unit, but it seems we're the first to see it.

To cross-check C-TAP, developed for FLIR, against the UFODAP algorithms for a ``regular'' camera, C-TAP was also applied to the 1716 UFODAP videos. It found $\ge$~1 valid triggers using its own logic in 85\% of the videos.

The C-TAP procedure consists of pixel-by-pixel subtraction of each frame from the preceding frame, similar to what has been done for bubble chamber imagery from direct dark matter searches~\cite{Szydagis:2010zz}. One single background~image for subtraction could not be generated, as the lighting and weather conditions were constantly changing. Furthermore, one of the goals was to~capture~craft in motion, making it necessary to default to using only the preceding~frame for comparison. This method is sensitive to objects either hotter/colder~than their environment. Either object type causes both negative and positive~values in subtraction, alternating, during a crossing of the field of view of one of the cameras. The best processing time of $\sim$1.5:1h~was achieved with GPUs.

After calculating the average in the pixel differences across an entire hour for each of the FLIRs, C-TAP sets a threshold 3-5$\sigma$ above this floor, unique to each video. It is adjusted depending on the nature of the noise, such~as~a severe non-Gaussian upward tail, typical of the older (noisier) camera units, necessitating setting a threshold toward the upper end of this range to avoid a high false-positive rate. A difference was discovered between camera noise and what appeared to be real/physical, moving objects based on the standard deviation $\sigma_a$ in six values: the greatest pixel differences in RGB, positive and negative. $M = max ( | p_f - p_i | )$, where~$p$ denotes an 8-bit (0-255) pixel value, and $M$ denotes the maximum negative difference (\textit{i.e.},~the minimum) or~the maximum positive difference. A lowercase or an uppercase subscript indicates minimum or maximum, respectively. The relevant Equations~(\ref{eqn:1})~follow:

\begin{equation}
\begin{split}
    \sigma_a = \sqrt{\Delta R^2+\Delta G^2+\Delta B^2+\Delta r^2+\Delta g^2+\Delta b^2} \\ \mathrm{where~\textit{e.g.}} \\
    \Delta R = M_R - \mu_a~\mathrm{and}~\Delta r = M_r - \mu_a \\
    \mathrm{and~similarly~for~G,~g,~B,~and~b.~The~mean~is~defined~as} \\
    \mu_a = ( M_R + M_G + M_B + M_r + M_g + M_b ) / 6
\vspace{-0pt}
\end{split}
\label{eqn:1}
\end{equation}

\begin{figure}[ht]
\centering
\includegraphics[width=1\textwidth,clip]{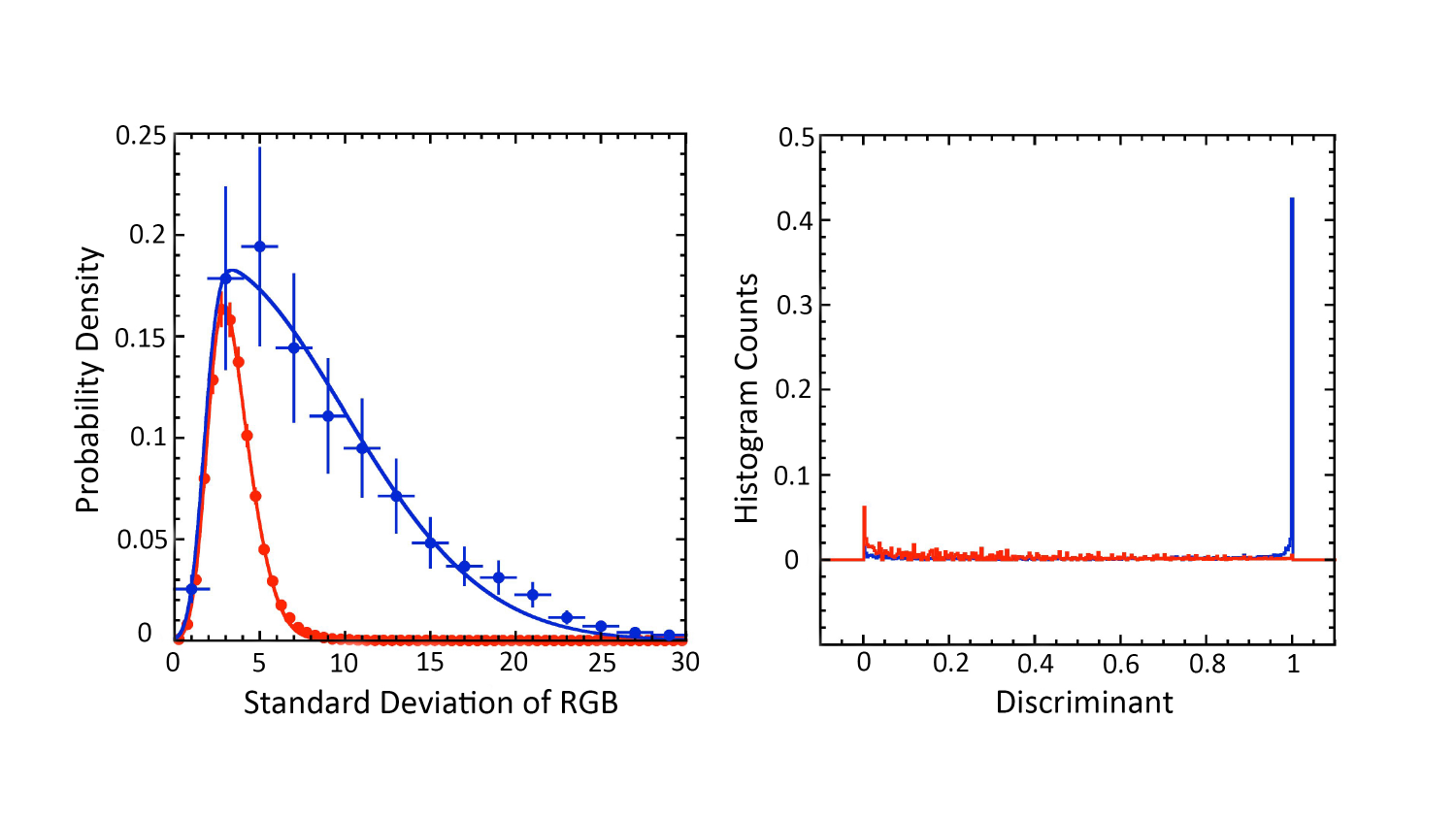}
\vspace{-48pt}
\caption{Left: Histograms of the standard deviations $\sigma_a$ from ``good'' (signal) and ``bad'' (background) events in blue and red, respectively, with skew-Gaussian fits to each. The y errors are $\sqrt{N}$, based on the number of events in each bin, while x errors are bin widths. One sample hour of FLIR video data was used, chosen for its mixture of clear and cloudy conditions and lighter mixed with darker conditions (dusk), and medium camera quality. This plot is a final asymptotic result from many iterations of human-eye scans at different thresholds to separate signal from backgrounds. Right: The classification-score output of a na\"ive Bayesian classifier based upon the skew Gaussians at left, for separating signal from backgrounds. 0.5 was the default value for separation (as is common for neural networks).}
\vspace{-8pt}
\label{Fig3}
\end{figure}

\noindent
Noise tends to increase or decrease the pixel values together across the 3 color values red, green, and blue, leading to a smaller $\sigma_a$. Meanwhile, true external events tended to have a greater value (Figure~\ref{Fig3}). As microbolometers simply detect IR photons, yielding only one measurement (of intensities), this again implies glitches existing in the internal false-color-rendering software.

In order to determine which imagery seemed the result of physical events external to the camera, for effectively defining a control data set for C-TAP training, initial C-TAP post-processing triggers resulting from either single-pixel or single-frame anomalies were rejected as false. Furthermore, volunteer video reviewers would (slowly and fully) watch videos to look out for discrete clusters of pixels traversing the screen, in two or more frames, taking either a linear or curved path. Different individuals would review the same videos as a cross-check, to reduce human bias. C-TAP parameter values for~the different steps of its processes summarized here would be updated until an optimum balance was found, maximizing the acceptance of events which appeared to be legitimate while minimizing the acceptance of those which appeared to be background noise only, with no clear object in view. (See Appendix 2.)

The process of human validation for tuning C-TAP results was iterative. A team member would view an entire video, making a list of frames of possible interest. Both the same and a different teammate would verify C-TAP agreed on these frames. Both would go back to confirm the objects caught by C-TAP but not by eye originally were visible, with the individuals involved logging both false positives as well as false negatives, clear objects (any kind) missed. C-TAP settings would be adjusted to reduce both types of events. To reduce fatigue, the videos would be shortened with each viewing to only frames still deemed of interest. On average, this method tripled the number of detections.

Classification of objects deemed legitimate by the combination of human QA and automation will be a second C-TAP module. Wing flapping for~birds and FAA records for airplanes, \textit{e.g.}, can help identify known phenomena for training. After it is complete, C-TAP will be made open source (on GitHub).

\subsection{Radiation Detection}

Some documented UAP encounters involve apparent harm from ionizing radiation,~\textit{e.g.}~the Cash-Landrum~\cite{CashLandrum} and Rendlesham Forest~\cite{Rendlesham} incidents. One of the benefits of working with radiation detection is that it is typically easier to analyze compared to images, as radiation detection can be as simple as lists of times and energies. As far as what is knowable from public sources, there are also no known air- or spacecraft in current operation that produce measurable quantities of radiation, at least not at large distances.

Terrestrial and cosmic sources exist and vary in intensity by location, so a coincidence measurement, combining radiation sensing with \textit{e.g.}~a camera ambiguity, is key for identifying potential multi-channel events of interest. An event superficially anomalous in terms of its energy may be a solar or galactic cosmic ray, or be extra-galactic, originating from a distant supermassive black hole or a supernova. Cosmic rays have been measured beyond 10$^{14}$~MeV~\cite{Albuquerque_2010}, and gammas $>$~10$^8$~MeV, traced in the latter case to the Crab pulsar~\cite{ANG_NER_2023}. The higher the energy the lower the flux, so the lower the probability of detection.

\begin{figure}[ht]
\centering
\includegraphics[width=0.9\textwidth,clip]{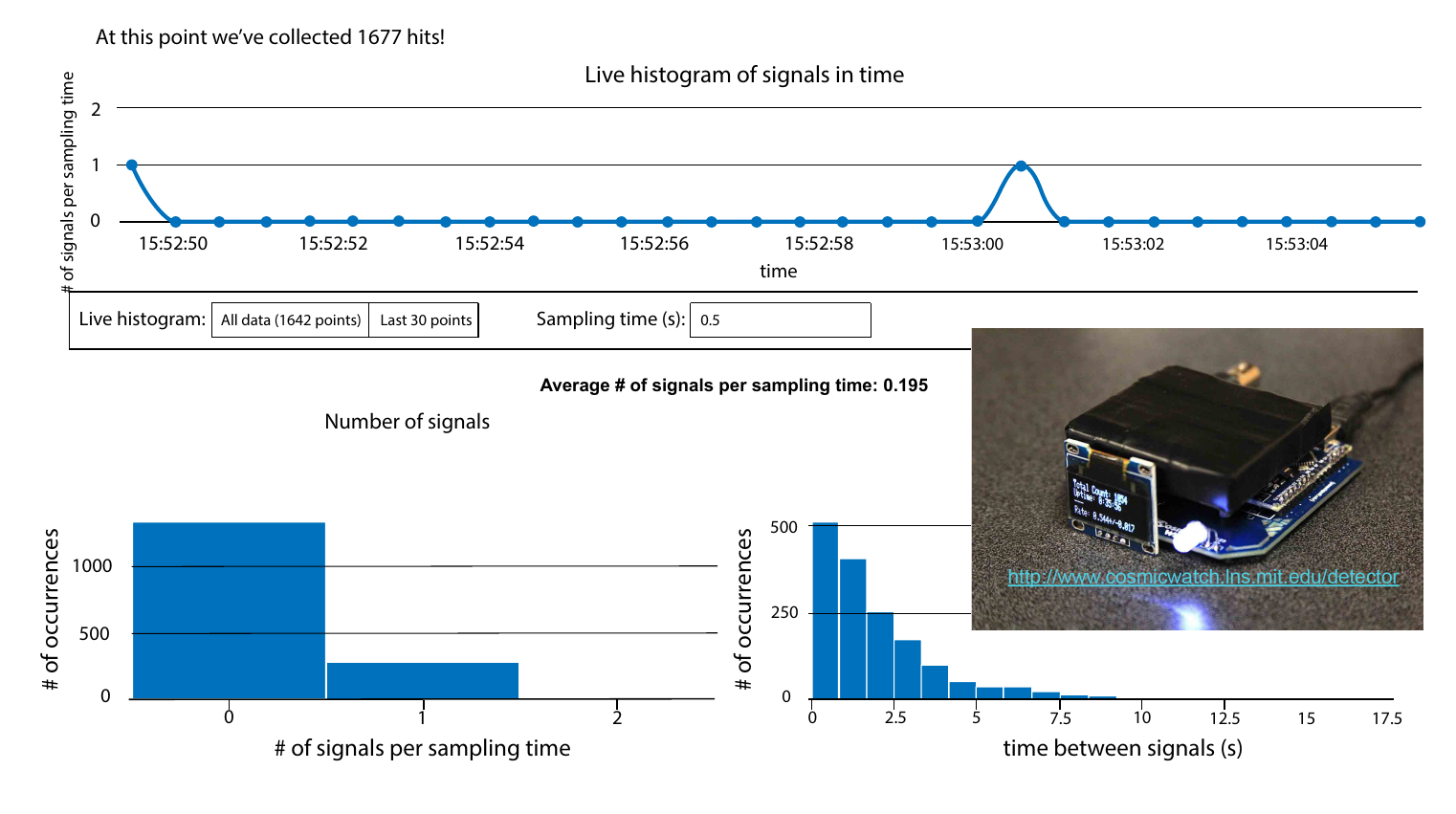}
\vspace{-15pt}
\caption{Screenshot of the live Cosmic Watch software (site). Inset: Unit without a cover.}
\vspace{-10pt}
\label{Fig4}
\end{figure}

UAPx uses the Cosmic Watch developed by MIT~\cite{Axani_2017,Axani_2018}, intended primarily for muon detection, but capable of detecting various different interaction types (although without particle ID or direction). The Cosmic Watch has the advantage of measuring both count rate and energy, not simply the former as in the case of the older Geiger-M$\ddot{\rm u}$ller counter. It is also smaller and trivial to deploy, coming with existing software (Figure~\ref{Fig4}) for USB data taking.~The device itself consists of 5$\times$5$\times$1 cm of plastic scintillator instrumented with~a SiPM (silicon photo-multiplier) operated by an Arduino Nano (Figure~\ref{Fig4},~inset). Every unit constructed will have variations in threshold and efficiency, leading to distinctive default rates. The exact unit used at Laguna Beach,~CA was calibrated both by just measuring the ambient radiation environment at UAlbany SUNY over the course of the year preceding the expedition, as well as with multiple alpha, beta, gamma, and neutron sources to ensure robust energy reconstruction ($dE/dx$, not energy, for non-stopped particles).

While the source calibrations were focused on deposited energy, the rate of cosmic rays is well known as a function of both the altitude and latitude and well studied utilizing Cosmic Watches~\cite{axani2019physics}. Particle flux will increase with~the former (up to a point) with less atmosphere for particles to traverse, and also increase with the latter, closer to the geomagnetic poles (due to the charged-particle component). A count rate vs.~energy for UAlbany (compared to~the spectrum from UAPx's first expedition) can be located in Section 5. Unique, newer units purchased after the expedition have had additional calibrations through cross-country drives, airplane trips, and underground lab visits.

\subsection{Future Instruments}

Next, we briefly describe instrumentation planned for future use and why.

\underline{(A.) Improved FLIR cameras (one already acquired: see Appendix 2)}

Lightweight models usable via smartphone can be slewed on a gimbal to point in the same direction as the UFODAP, which should be able to share angular information as it follows a target. Claims exist of anomalous objects in the sky, observable in one wavelength range but not in another (\textit{e.g.}, only IR-detectable~\cite{Brothers} or in observable light only). These imply unknown natural phenomena and/or signature management.

\underline{(B.) Ultraviolet (UV) imaging (one camera already acquired)}

UV is an unexplored, new realm for UAP research. Adding sensitivity to it is motivated by claims of Doppler blueshift from exotic propulsion.

\underline{(C.) Electromagnetic field sensing}

Many reports associate UAP with electromagnetic effects, such as vehicle disruption and deactivation, and other, electronic equipment failures~\cite{catoe1969ufos,Falla:1979,Rodeghier:1981,McCampbell,PowellLittle, Magnet, MADAR}. We'll thus employ off-the-shelf 3D EMF / trifield meters, plus smartphone magnetometer apps that use a phone's existing hardware. These have the advantages of low cost, portability, and ease of usage. A quick calibration of one (Appreciate Studios), moving the iPhone running~the~app away from a strong rare-earth magnet in discrete steps along a ruler, yielded a power-law exponent of -2.74 $\pm$ 0.09 for the magnitude of the measured B-field versus distance, not far from -3 (theoretical value for a perfect dipole).

\underline{(D.) Magnetic gradiometer}

A B-field gradiometer will supplement (C.) but also detect metal remotely. The planned device should be able to sense a passing car as far as 30m away.

\underline{(E.) Piezo-electric acoustic sensors (two acquired)}

Piezos will be deployed to seek anomalies from the infrasound to the ultrasound range~\cite{PowellLittle}, which could be caused by air disruption due to hypersonic travel~\cite{Mead_2023}. Infrasound may also induce hallucinations, so that it is a plausible explanation for the more unusual claims from UAP eyewitnesses~\cite{french2009haunt,GermanAnoNewest,UK}. UAlbany owns piezos previously useful for (dark-matter-seeking) superheated droplet detectors, though these listened mainly for ultrasound~\cite{Aubin_2008}. UAPx~has also attempted to access data from existing global infrasound networks used to ensure compliance with nuclear test-ban treaties, though without success.

As always, to confirm if any ambiguous findings can robustly be claimed to be anomalies, coincidence windows with other sensors remain key, especially as calibration and background characterization in a noisy environment~(sky, sea, ground) will be challenging even given piezo frequency analysis (FFT).

\section{Laguna/Catalina Expedition: Successes and Lessons Learned}

While several observations were at least initially intriguing, the primary purpose of the initial outing, in retrospect, was an effective field testing of all the equipment and analysis techniques, even though it had novel facets, such as a Cosmic Watch for coincident particle detection, novel software (C-TAP, similar to what was presented in \cite{Cloete} but for FLIR), and the first time UAP seekers capitalized on Doppler radar (Sec.~5.1.1). Lessons learned included:

\begin{itemize}
\vspace{-2pt}
\item {The UFODAP hardware should be of sufficiently high quality to serve as scientific instrumentation, but its software is not reliable for object tracking and identification, nor is the software capable of accessing the MSDAU (Multi-Sensor Data Acquisition Unit), the name given to the non-optical sensor collection, and as a result no ancillary data, such as GPS location and ADS-B exchange, were recorded.}
\vspace{-2pt}
\item {Tracking just aircraft, via transponders, is not sufficient. It is also~necessary to have apps showing maps of satellites (especially Starlink, often mistaken for UAP), all known rocket launches, and the ISS' trajectory. (See Appendix 1 for a quantitative study of our ISS observation.)}
\vspace{-2pt}
\item {Multiple identical cameras are still a necessity, even if others can supplement the UFODAP like FLIR, all with sufficient (12-hour+) battery backups.}
\vspace{-2pt}
\item {Having many FLIR cameras was not as beneficial as originally expected. They generated a large quantity of data which were not of the highest quality and challenging to analyze within a reasonable timeframe given limited personpower / CPUs. Instead, fewer, more modern, and better calibrated units would be optimal, as per Section 3.4(A.) Any connector must be RF-shielded (with metal) against stray EM noise.}
\vspace{-2pt}
\item All clocks must be synchronized, ideally at the sub-sec.~level, and device positions recorded using a combination of GPS with a laser range finder. Working with a film crew, necessary for our seed funding, created distractions which led to these, and other, critical steps being~neglected. One can imagine that the researchers who have at times labored under similar conditions have been affected in similar ways~\cite{CKelleher,JTLacatski}.
\vspace{-2pt}
\end{itemize}

Lastly, more than two Cosmic Watches running in coincidence mode allow for a reduced background rate as well as rudimentary directionality. Alternatively, separation of a pair outside the typical diameter of cosmic-ray~showers results in a greatly reduced background, permitting any anomalies to be more obvious. It is important to note, however, that due to secondary particle production from GeV-TeV muons, shielding will increase the background rate.
\vspace{-15pt}
\section{Example Observations and Preliminary Results}
\vspace{-5pt}
\subsection{UFODAP: Its Titular Ambiguity}

\begin{figure}[ht!]
\centering
\begin{subfigure}
    \centering\includegraphics[width=1.60\textwidth,clip]{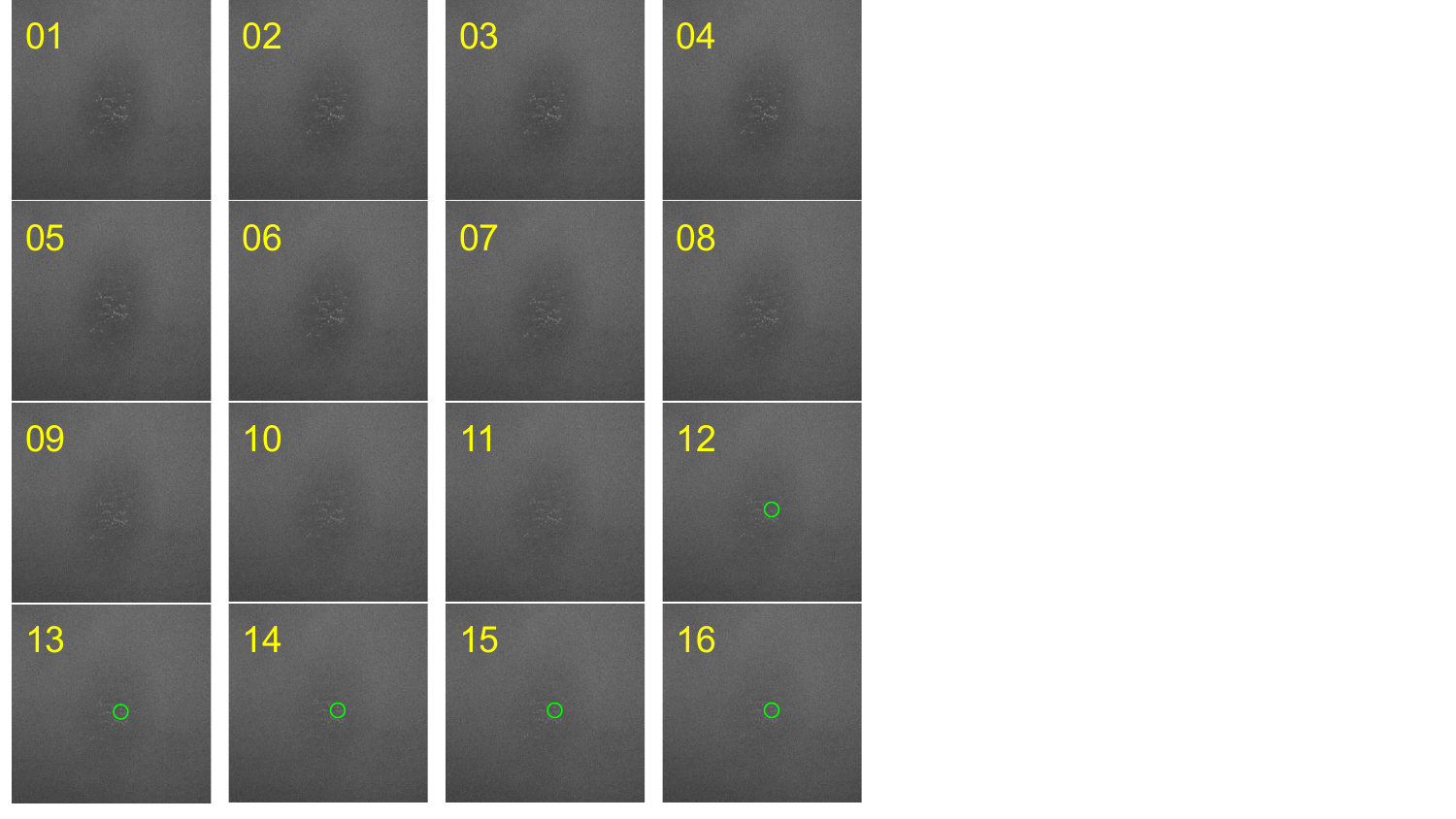}
    \end{subfigure}
    \begin{subfigure}
    \centering\includegraphics[width=1\textwidth,clip]{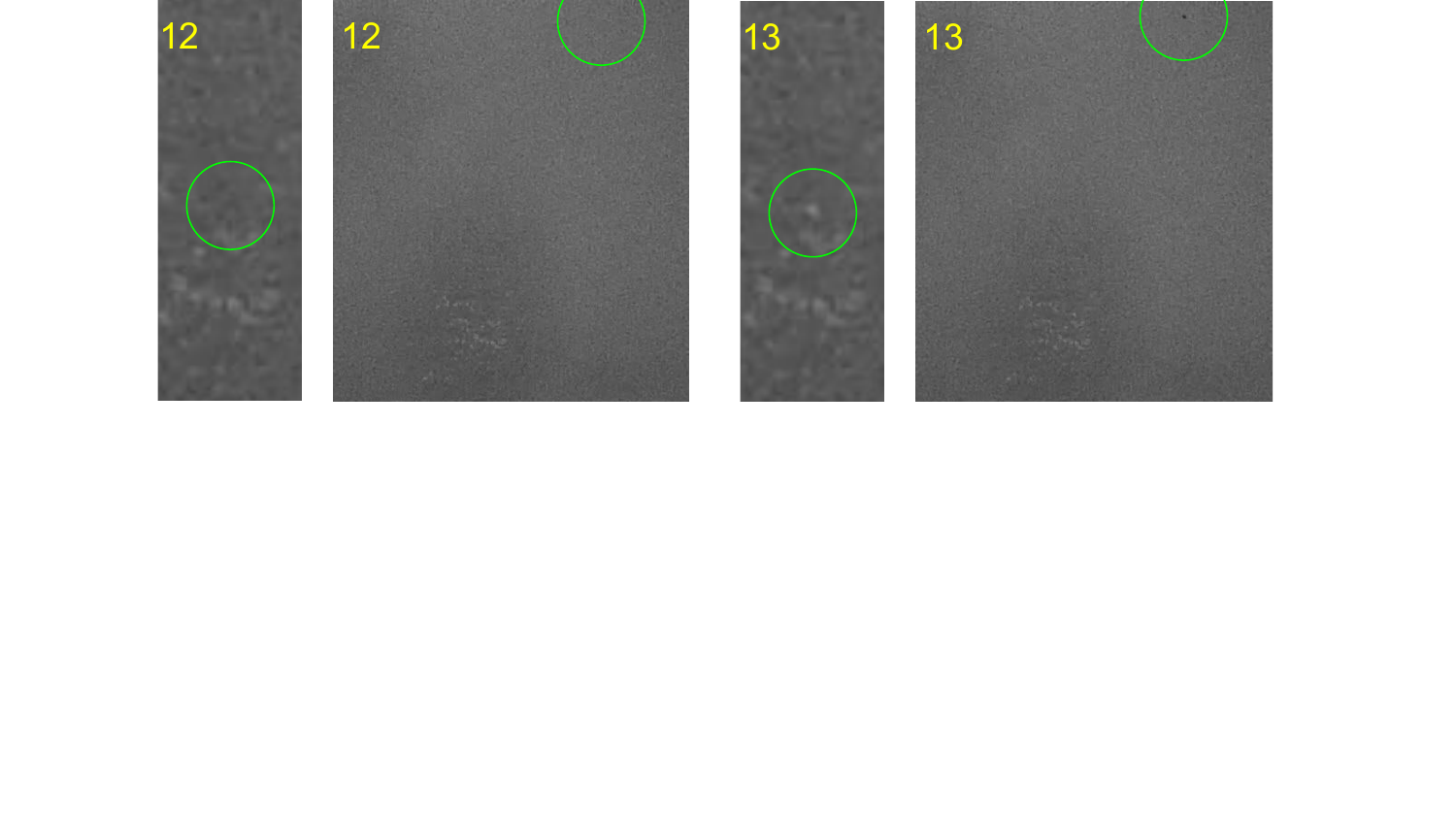}
     \end{subfigure}
\vspace{-130pt}
\caption{Top: A zoom-in on the dark spot from 3:59:24am with green circles indicating~the appearance of a new white dot between frames 12 \& 13. Bottom: Further zoom on frames 12-3 only for greater ease of seeing its appearance, concurrent with a black dot emerging. The behavior of our dots is similar to that of the historical transients discovered in \cite{solano2023bright,Villarroel_2021}.}
\vspace{-0pt}
\label{Fig5}
\end{figure}

The UFODAP system ultimately made a detection that remains the expedition’s most intriguing ambiguity. Multiple camera ambiguities (multiple videos close by in time on the minute scale) were discovered, but neither live via visual observation by a team member, nor in a systematic review of the recordings by person or software. These ambiguities were discovered through a systematic review of the Cosmic Watch data, because they appear to be associated, at least temporally, with the highest-energy event measured in the Cosmic Watch, by itself not necessarily anomalous, as explained in Sections 3.3 and 5.2. While there were potentially corroborative same-camera videos, no second camera made an observation which could have settled the crucial question of: internal to the UFODAP or external? The FLIR cameras, some pointed in the same direction, were not active at the time due to the power needs of film equipment. This is a timeline for Friday~07/16/21, in PDT:
\vspace{-2pt}
\begin{itemize}
    \item 3:50:14am. A ``blank'' video $i.e.$ without an obvious trigger condition
    \vspace{-3pt}
    \item 3:57:16. A diffuse dark spot appears at upper right. Its appearance may have triggered the UFODAP. It has no well-defined edge (Appendix 3).
    \vspace{-20pt}
    \item 3:57:27. The spot remains; the camera slews, chasing an insect separate from it. During rotation, white dots that~turn into black streaks appear, emanating from this spot. Note that the spot is not visible in the later frames where the background sky is well-illuminated. (See Appendix~3.)
    \vspace{-20pt}
    \item 3:59:24. The camera is again stationary and the dots are visible within the spot for the entire video's length, though the spot vanishes (lightens to match the surrounding gray cloud color) and a new white~dot~appears in frame 13 where the spot was located, along with a new black~dot,~in~a different region. Frames from every video are in Appendix 3 (at full res and zoomable) but those from this one most important video in Fig.~\ref{Fig5}. The transient nature of our dots is not inconsistent with \cite{Villarroel_2021,solano2023bright}, and they may be part of a significant larger background of objects~\cite{villarroel2022background}.
    \vspace{-3pt}
    \item 04:00:13am. The first ``normal'' video after the incident. This video~and all of those above were quite short (a few seconds each at the most).
\end{itemize}

\textcolor{white}{.}
\vspace{-0pt}

\begin{figure}[th!]
\centering
\includegraphics[width=1\textwidth,clip]{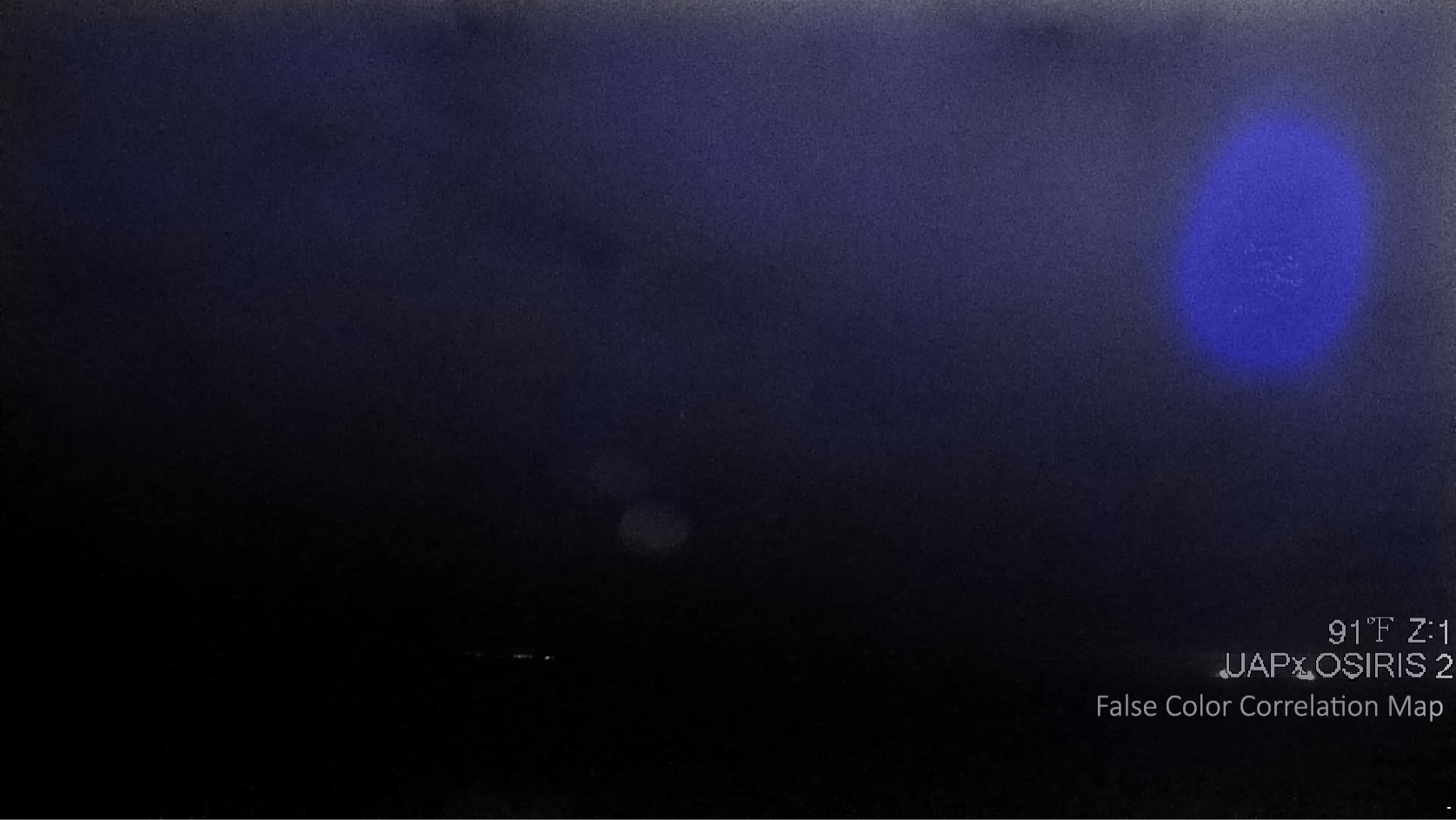}
\vspace{-10pt}
\caption{A false-color (blue) correlation map of pixel intensity behavior over the 12 (out of 16) frames of the dark spot vanishing. There exists some correlation between the behavior of the cloud cover and the behavior of the dark spot, implying it is an atmospheric effect (the faint blue may also be a fly departure trail). Its disappearance could have been due to a lower cloud level condensing across the sky, but then the white dots remain unexplained.}
\vspace{-0pt}
\label{Fig6}
\end{figure}

We evaluate ten prosaic (``null'') hypotheses put forth both inside/outside our group. Our list is an attempt at completeness, for ruling out or in, even though that is never fully practical. (However, in this case and in others,~UAP cannot just ``be anything,'' a common phrase used in debunking.)
\vspace{-0pt}
\begin{enumerate}
    \item Fall-streak hole from aircraft, with the white dots the aircraft, or noise (or:~natural cloud formation, Fig.~\ref{Fig6}, least improbable by elimination)
    \item A star field or seagull flock producing white dots, as viewed through a hole in the clouds, the dark spot, either a natural or fall-streak hole
    \item Water drop evaporating, evident as a spot slowly decreasing in area
    \item Fly on the protective dome leaving, evident as a spot suddenly gone
    \item Cosmic-ray shower, where ionization lights up CMOS pixel clusters
    \item Meteor breaking up or meteor shower: fragments of a single meteor or multiple meteors showing up as dots, and making a hole in the clouds
    \item Camera noise in the dark environment, possibly combined with a residual effect (algorithmic artifact) from the camera rotation earlier -- this explains white dots and streaks, but not the dark spot, or its fade-out.
    \item Resetting of camera levels for a lightening sky, which should manifest as a change in pixel intensities across an entire image (dynamic scaling)
    \item Military testing tied to a nearby base and/or training area of operation, where the white dots are \textit{e.g.}~a drone cluster, creating a hole (1)
    \item The reflection of lights in the camera dome from the nearby cities leading to white dots, inside of a water drop for instance (Hypothesis 3).
\end{enumerate}
\vspace{-4pt}
Note that having two or more cameras (\textit{e.g.}, the fisheye or one active FLIR) could have eliminated hypotheses 3, 4, 7, 8, and 10, leaving only five (external) explanations and greatly facilitating assessment of observations.
\vspace{-14pt}
\begin{figure}[th!]
\centering
\includegraphics[width=0.93\textwidth,clip]{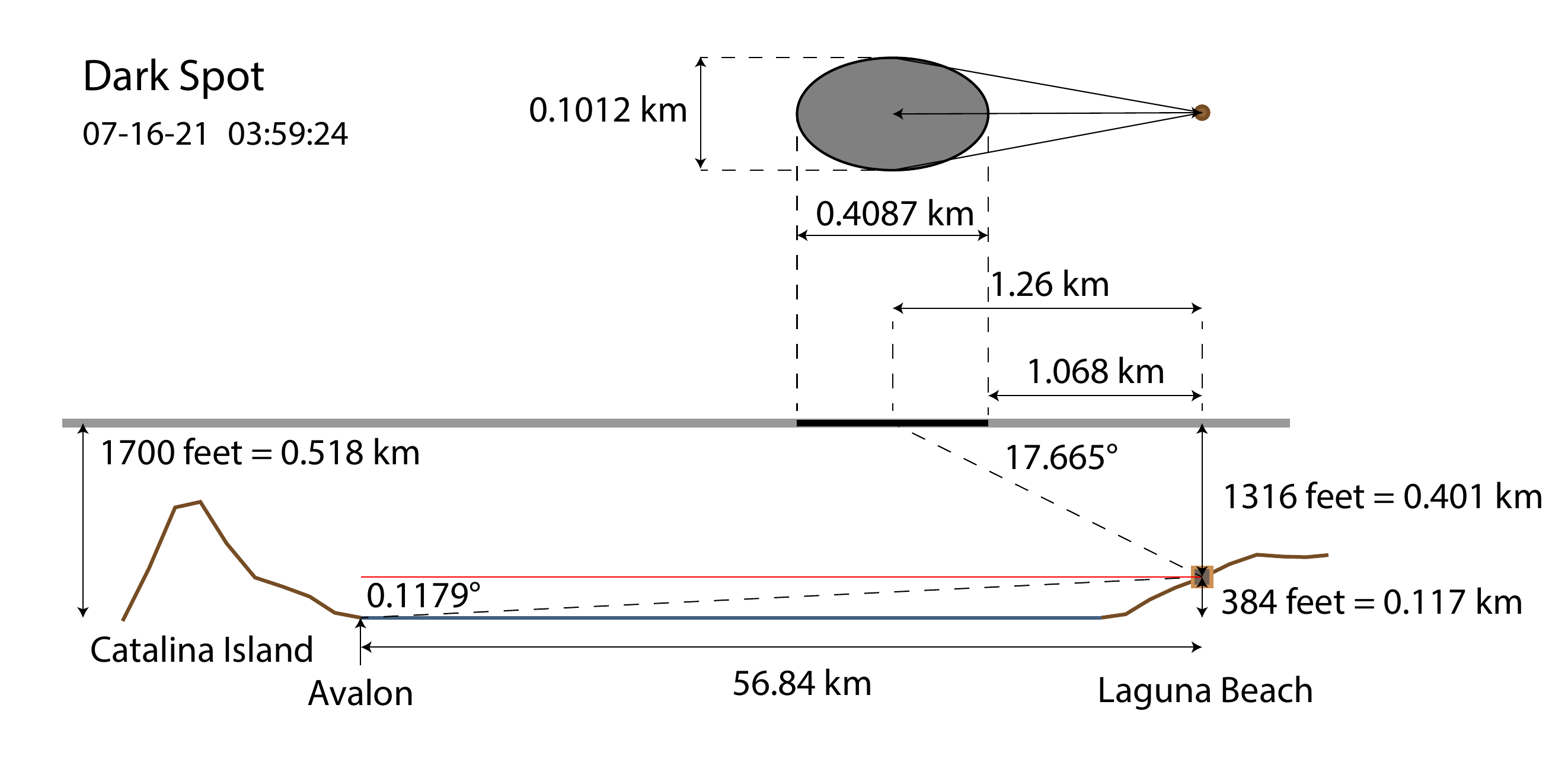}
\vspace{-20pt}
\caption{An illustration of the geometry of the dark spot assuming that it was a physical, elliptical hole in the clouds within the cloud ceiling at the time. If the white dots~were~real objects at/above the ceiling, they were each $\gtrsim$4m wide, using equations in Appendix 1.}
\vspace{-4pt}
\label{Fig7}
\end{figure}

The fall-streak hole hypothesis (1) explains a dark spot, but not the~white dots (bright clusters of pixels) although those are potentially explained by the camera noise hypothesis (7) or aircraft. By itself (1) is incomplete:~no~aircraft could be identified as being in the area at the time, to cause a fall-streak hole. Moreover, the atmospheric conditions were far beyond what is necessary for a naturally-occurring fall-streak hole (\textit{i.e.},~it was not sufficiently cold not even at the altitude of the cloud layer for supercooled water to exist then freeze). If this were a natural hole between clouds, then the star field hypothesis (2) would be possible, but it is not likely due to the fact the pattern of dots is~far too dense (considering LA's light pollution) and does not match the known local sky (\textit{cf.}~0.2 dots/deg$^2$ to 2000/40000=0.05 stars/deg$^2$ for a dark sky). The dots are too bright to be high-altitude seagulls instead, and it~would be improbable for them to be clustered in the hole, even accounting for flocking.

Unlike the fall-streak hole hypothesis (1), the star field or flock of seagulls

\noindent
hypothesis (2) cannot resolve the rapid lightening of the dark region,~since~the distance and size, as illustrated in Figure~\ref{Fig7}, would imply clouds moving~on the order of 100~m/s, although the behavior of the dark spot, correlated with that of the overcast sky (Figure~\ref{Fig6}), suggests that the spot was occluded by condensation of a lower level of humidity across the visible sky. Even if the atmospheric conditions were correct, this fails to address the dots remaining, if they are camera noise (not likely to be the case, as we quantify later).

The next two hypotheses involve something close to the camera instead of a physical event detected in the clouds, so they are sequential in our list. The dark spot fades away much too quickly to be water (3) evaporating naturally, especially since evaporation occurs at a liquid droplet's surface. That means the dark spot would slowly decrease in size as evaporation occurred. This is not what happens---the spot lightens uniformly. Moreover, there was no rain at the time and the temperature was well above the dew point. The~house~on whose roof the UFODAP had been deployed was 120 m above sea level and too far inland (0.45 km) for sea spray to be a possibility. If this were a drop of dew, it should've had a mass $\approx$50~mg so we can estimate the power needed to evaporate this water mass at a rate matching that of the spot fading:
\vspace{-0pt}
\begin{equation}
Q = m L_v = (0.05~\mathrm{g})(2260~\mathrm{J/g}) = 113~\mathrm{J}.~P = (113~\mathrm{J})/(0.6~\mathrm{s}) = 188~\mathrm{W}.
\vspace{-0pt}
\label{eqn:2}
\end{equation}
\noindent
Even given an order of magnitude error such sudden power is unreasonable.

The fly explanation (4) is similar, but is more plausible. However, a~fly's presence on the outer dome does not explain why the first 30 or so frames of the 3:57:27~am video (Figure~\ref{Fig15}) show the dark spot, while the later frames do not, with the spot reappearing for the 3:59:24~am video (in Figure~\ref{Fig16}).~The fly hypothesis also does not explain why the behavior of the lightening dark spot is correlated with the behavior of the cloudy sky (Figure~\ref{Fig6}). Only~a~fly taking a specific, slow departure path could exhibit such a correlation. Additionally, the fly hypothesis requires something like camera noise (7) for the white dots; the fly would provide the dark spot, with thermal/shot noise within it.

Cosmic rays (hypothesis 5) seem even more natural, given it was a Cosmic Watch detection which led to the discovery of the videos in question and the fact that CCDs are known for particle detection via ionization.~They~can~be employed as dark matter detectors~\cite{Chavarria_2016}. However, the localization is not natural, especially since camera lenses are only able to focus observable light~(not gamma rays nor cosmic rays, of course). A cosmic ray shower should have~affected most of the field in the CMOS, not CCD in our case---not just a particular region. Moreover, the cosmic ray hypothesis doesn't address the dark region in general, and~the improbability of all of the dots to be within it.

Another explanation (6) involving an external, astronomical origin would be meteor(s). It fails as well, since no dramatic meteors were reported in the area at the time, neither one breaking up, nor a shower of multiple, and such non-streaking (static) behavior is not common, requiring a head-on view.

The noise hypothesis (7) is promising, not only for the white dots, but the dark spot itself, as the latter's darkness (still requiring its own explanation) may have caused the former, but this type of noise was only observed during lens motion, never in stationary cases, like Fig.~\ref{Fig5} (+\ref{Fig16}). All crew were asleep at the time and accounted for, and the roof was accessible by scissor lift only, making it implausible for a passerby to shake the system inadvertently. Most importantly, the brightness of the white pixels is not Poisson-distributed, as examples of camera noise superficially (visually) the same are (see Table~\ref{Tab1}).

\begin{table}[h]
\centering
\begin{tabular}{||c c c c c||} 
 \hline
  & Case A & Case B & Case C & Case D \\ [0.5ex] 
 \hline\hline
 $\mu$ & 82 & 89 & 65 & 66 \\ 
 \hline
$\sigma^2$ & 51 & 98 & 21 & 40 \\
 \hline
\end{tabular}
\caption{The means and variances of sets of 8-bit pixel values. Case A is the dark spot, B is the same area but within an image without the spot, C is a control area from the same image as A, and D is for the same control area as B for an image without the dark spot.}
\vspace{-10pt}
\label{Tab1}
\end{table}

\noindent
The table shows the variance was 51 for a mean of 82, as reported on a black-and-white scale of 0-255. This is sub-Poissonian by nearly 40\%. For a Poisson distribution, as expected for shot noise, the mean ($\mu$) and the variance ($\sigma^2$) should be equal. In a visual example of (different) white~dots seen at another day and time utilized as controls, choosing dots with a mean comparable to that of the white dots in question, the resultant $\sigma^2$ was only 10\% off from the Poissonian expectation (B). C and D demonstrate that image areas without noise exhibited sub-Poissonian $\sigma^2$s, by similar or greater margins.

The closely related resetting of camera parameters hypothesis (8) is disproven by a pixel intensity behavior correlation map for the ``closure'' (Fig.~\ref{Fig6}).

It is not realistic to address the military hypothesis (9), without also~hypothesizing details, due to classification.

The light reflection hypothesis (10) must be carefully studied, even though it cannot explain why this effect was not observed on other nights, nor can~it address the dark spot and its slight correlation with the overcast sky.

Most significantly, if the upper-right dark spot is not within the sky, but something on the dome, it does not explain the multiple corroborative videos close in time, which sometimes included camera rotation, both with/without significant spot motion, as well as the possibly corroborative radiation data.

While no single hypothesis fits all the facts, many have still been tested: water drops have been placed and allowed to evaporate, and objects approximating insects, for testing of the water droplet (3) and insect (4) hypotheses (Appendix 3). Exposure to many distinct radioactive calibration~sources~in~a lab (in Appendix 3) probed the cosmic-ray shower hypothesis (5), but~a~multi-week dark-box test, to probe the cosmic-ray shower (5) and camera noise (7) hypotheses together, is yet to be executed, for a near-future paper. LEDs~and flashlights shined into water droplets are some of the ways in which one can directly probe the light reflections hypothesis (10). Tests~like~these~can~be~important for establishing the baseline behavior of our equipment in the face of common phenomena. Most importantly, the pixel intensities mean, intensity width, spatial spread, and other calculations will remedy the lack of quantitative analyses for some of the hypotheses dismissed qualitatively,~for~now. Analyses of (publicly-available) radar data seeking any evidence for / against objects being present in the sky and thorough background from \cite{RADAR} are next.

\subsubsection{Doppler Weather Radar}

In an effort to answer the crucial question of whether an internal (to~the UFODAP) or external phenomenon had been observed, past data from The Next Generation Weather Radar (NEXRAD, part of NOAA) network covering the contiguous U.S.~were accessed. (Three sites of WSR-88D installations had the potential to have measurements for our area of interest.) To the best of the authors' knowledge, this is the first attempt to corroborate UAP camera detection by analyzing data from this data source, in a scientific paper. While FAA radar has been used, $e.g.$~by Robert Powell (author of \cite{PowellUFOs}) in the Stephenville, TX case, use of public weather radar data, with control cases, is novel. Types of data recorded by the NEXRAD radar system include:
\vspace{-6pt}
\begin{enumerate}
    \item Reflectivity
    \vspace{-5pt}
    \item Radial velocity
    \vspace{-5pt}
    \item Spectral width
    \vspace{-5pt}
    \item Correlation coefficient
\end{enumerate}
\vspace{-6pt}
(1-3) are measured from one beam to the next in ``moments.'' (4) is measured from the contrast between the vertically- / horizontally-oriented radar~pulses. Reflectivity was of particular interest, specifically clutter reflectivity, defined later. Reflectivity (1) is amount of transmitted power reflected back to the radar station. For NEXRAD, this is defined with respect to a reference power level of 1 mW, although the reference utilized in radar is typically expressed in decibels (dB). Velocity (2) means average toward~/~away from the station. Spectral width (3) corresponds with directions of velocity vectors, for detections in a ``radar pixel,'' a unit of volume through which a beam~passes.~Low implies high consistency in direction; high, greater randomness. Correlation coefficient (4) is a value describing how uniform the shapes of observed features are, based on comparisons of the reflectivity returning back to the radar from its horizontal and its vertical pulses (\textit{e.g.}, a perfect circle would deliver a correlation coefficient of 1; in practice, rain or snow is known to approach a coefficient of 1, while birds and insects tend to produce values below 0.8). These variables can help determine characteristics of an object that may be detected: info on size, shape, and its state of motion relative to the station.

Clutter is defined as radar reflections which are deemed to be from ``non-meteorological events'' and filtered out (such as traffic or wind farm clutter). The NEXRAD data set saves these in a separate variable called clutter reflectivity (``clutter\_filter\_power\_removed''). Filtration of clutter is a complicated problem: assumptions have to be made. Typically, all detections with~no~velocity are filtered out, along with manual filtering for known sources of clutter. Different processes are implemented for different radar beam elevation angles. NEXRAD additionally uses a ``fuzzy logic'' algorithm for more intelligent filtration, permitting more true meteorological events to be logged~\cite{FuzzyLogic}.

Using NWS' (National Weather Service) NOAA glossary, we list known levels of non-clutter (weather) reflectivity. VIP (Video Integrator/Processor) contours it (in dBZ, unweighted unlike dB units for audio) at 6 levels:
\vspace{-8pt}
\begin{itemize}
    \item VIP 1 (18-30 dBZ) - Light precipitation
    \vspace{-13pt}
    \item VIP 2 (30-38 dBZ) - Light to moderate rain
    \vspace{-13pt}
    \item VIP 3 (38-44 dBZ) - Moderate to heavy rain
    \vspace{-13pt}
    \item VIP 4 (44-50 dBZ) - Heavy rain
    \vspace{-13pt}
    \item VIP 5 (50-57 dBZ) - Very heavy rain; hail possible
    \vspace{-13pt}
    \item VIP 6 ($>$ 57 dBZ) - Very heavy rain and hail; large hail possible
\end{itemize}
\vspace{-8pt}

If the dark spot was an atmospheric event, then based on camera location and angle it most probably occurred between 281.1$^{\circ}$ and 286.4$^{\circ}$ (WNW) from the measurement spot on the Californian coast (Figure~\ref{Fig7}). Radar returns~between 3:30-4:30 were explored starting with the closest station, KSOX. The hour was divided into 10s bins; one interesting time frame was identified,~possessing a detection within a pair~of rays determined from analyzing the dark spot's coordinates on the UFODAP videos, at 04:03:50 PDT.

\begin{figure}[ht!]
\centering
\includegraphics[width=1\textwidth,clip]{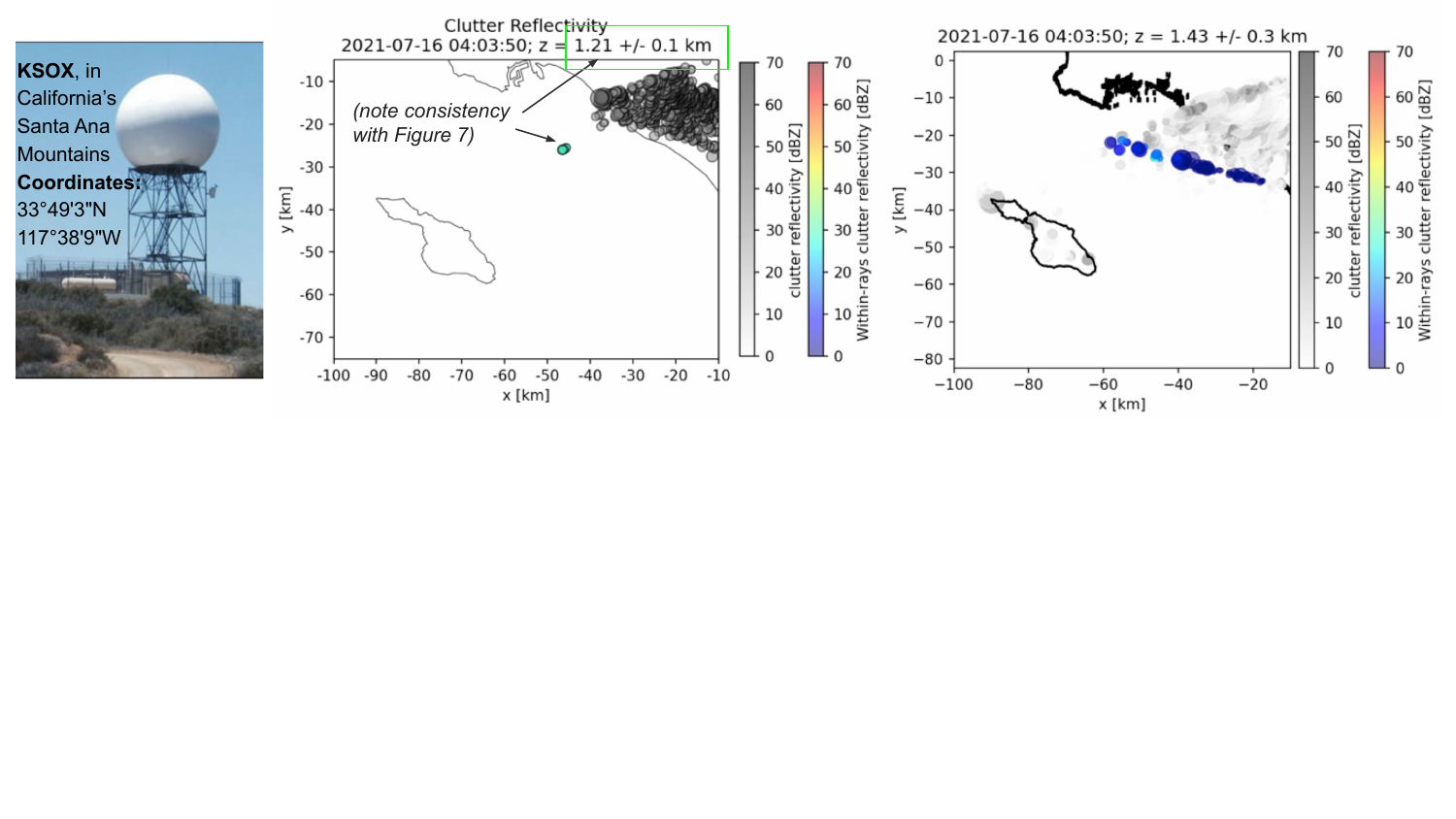}
\vspace{-130pt}
\caption{Left: KSOX. Middle: All of its detections within the dark-spot rays in color; all other detections in B\&W. Signals below 25~dBZ (default) were filtered out to ensure only strong ones are displayed. Our plot visualizes radar data differently than is traditional (10s bin rather than full sweep, showing circular markers scaling with reflectivity) to accentuate what is relevant for us. Full-sweep maps confirm the data are the same. $z$ indicates mean elevation for all detections in the bin. Right: No filter; there are many more detections~but most too low in intensity to be objects. A case could be made that the~(27-29~dBZ)~clutter event could be the densest portion of a then slow weather pattern that the radar algorithm designated as clutter. The following 3 days after this were a monsoon event in the region.}
\vspace{-0pt}
\label{Fig8}
\end{figure}

There were, in fact, three points inside of that pair of rays -- only slightly later than the times of interest, having values of 27~dBZ, 29~dBZ and 29~dBZ, large enough to not be likely ``sea” clutter (reflections from movements of the waves). See Figure~\ref{Fig8}.

The other two nearby stations, KNKX and KVTX, were also considered. The former was not sweeping the area at the same time, and the closest~sweep in time does not detect anything identified as clutter within the angles of~interest. The latter swept through at almost the same time and~at a relevantly close elevation, but did not have any detections. This may still be consistent with the KSOX observation, as radar pixels (\textit{range gates}) get larger the further away they are from the detector. Small objects may get washed out when sufficiently far from the station. Range gate is the area encompassed by one radar pixel; it becomes larger with increasing distance from the radar site, as the beam expands, moving away from the source. So resolution decreases with distance from the site. At 50 km resolution is 1 km$^2$ for a WSR-88D.
\vspace{-5pt}

\begin{figure}[hb!]
\centering
\includegraphics[width=0.99\textwidth,clip]{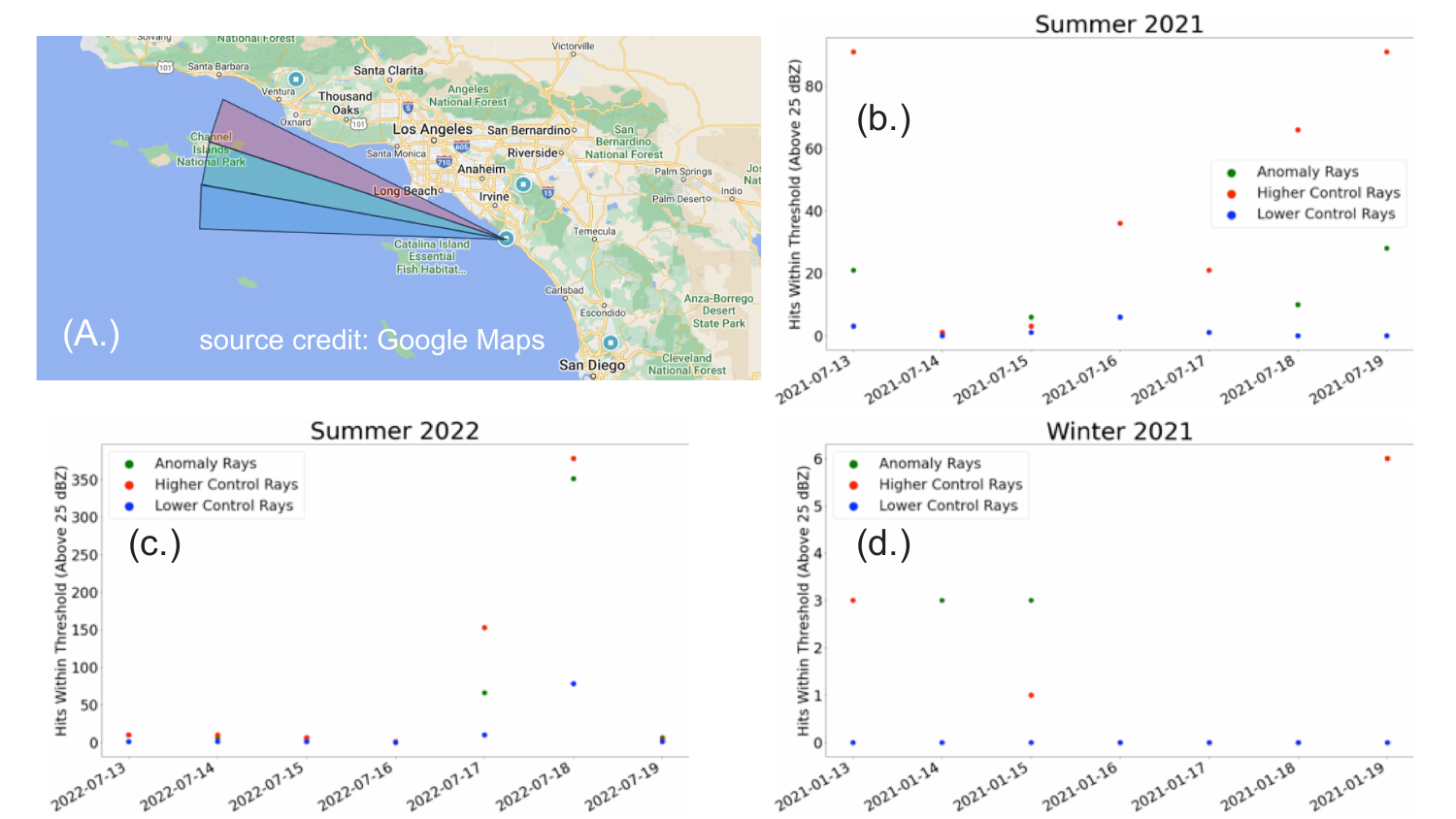}
\vspace{-2pt}
\caption{(A.) An approximate visualization for the ``ambiguity's rays'' (centered, green), Control 1 (red, higher) and Control 2 (blue, lower). (b.-d.)~are clutter reflectivity~hits~above threshold (25~dBZ) at different times, with (b.) corresponding with the UAPx expedition. Day-to-day variability seems lower for the lower-angle control rays, but with many more hits in summer than winter (note y-axis limits). This may suggest seasonal~variation.~Hits within the ``ambiguity'' rays are not particularly rare. Many more hits appear in the~higher-angle control rays, although this may be influenced by being closer to the shore.~The~large increase in hits in 2022 indicates year-to-year differences may be significant too, and Summer 2022 has large day-to-day variability (central-rays hits aren't more probable).}
\vspace{-0pt}
\label{Fig9}
\end{figure}

One can ask how interesting our KSOX frame is. Echoes from objects~like buildings and hills appear in almost all radar reflectivity images. This ground clutter generally appears within a radius of 40 km of a site, as a roughly~circular region patterned after the surface roughness of the terrain. Radar~returns from animals or aircraft are also not uncommon. The apparent intensity and aerial coverage of these features are partly dependent on radio propagation conditions, but they usually appear within 48 km of the station and produce reflectivity of $<$30 dBZ. So, the three points 27, 29, 29 may~be~considered at least as solid and sizeable of an airborne-clutter reflectivity event as birds, insects, or aircraft. The fact these are relatively high values, and are close~in time to the ambiguous videos and 1-2 high-energy particle detections is interesting. Figure~\ref{Fig9} addresses the potential explanation of a statistical anomaly.

No corresponding values exist for spectral width or radial velocity (data missing). Correlation coefficients ranged from 0.20-1.05: given their inconsistency, one cannot conclude much from them. Simple, preliminary statistical tests shown in Figure~\ref{Fig9} endeavor to answer how common high reflectivity~is within background data. We defined three sets of rays: the rays for the dark spot, control rays 1 adjacent but north of them and control rays 2 adjacent but south of them, always preserving the same angular width.~The~data~were obtained from other dates for the same hour (3:30-4:30am) at time scales of: $\pm$ 3 days, $\pm$ 6 months (winter vs. summer), and +1 year (data from the year before were found to not include a clutter reflectivity variable).
\vspace{-5pt}

The case can be made that it is remarkable that there are radar hits registered by the closest radar station at approximately the same time and place as the camera dark spot with white dots, but timing was indeed only approximate, and it fell outside of UAPx's timing resolution by a factor of~roughly four (Section 5.2). The altitude `$z$' (of 1.2~km) for the hits (Figure~\ref{Fig8}, center title) is a factor of three too large to match a hole in the cloud ceiling, unless more distant objects were being viewed through a hole in it. Outside of that context, however, the detections themselves remain statistically insignificant, according to Figure~\ref{Fig9}. Furthermore, we do not have enough evidence to definitively conclude that they are a part of the same effect(s)~/~event(s).

That leaves us with this provisional conclusion: that the 27-29 dBZ blips are not inconsistent with the videos and Cosmic Watch, \textit{i.e.}, neither corroborating nor contradicting a non-null hypothesis, or one of the ten involving aerial solids. But now a framework for this sort of analysis is in place, opening doors for comparisons at other locations where there are UAP-concurrent data, to aid in accumulation of corroborating data (\textit{e.g.}, on future trips).

\subsection{Radiation: Counts / Energies, and Systematic Uncertainties}

Multiple statistical outliers were identified within the 205,920~s of Cosmic Watch livetime (89,220 individual detections), as two types: high-rate~excursions (time-binning dependent) and higher energies (Appendix 4). In~the~former case, most appeared to be correlated with increased solar activity from the week of the expedition based upon data from the scientific instrument of ERNE (Energetic and Relativistic Nuclei and Electrons) for measuring energetic particles, deployed on the SOHO (Solar and Heliospheric Observatory) spacecraft, as well as data from particle detectors on GOES (Geostationary Operational~Environmental Satellite), or NASA geomagnetic storm alerts.

Cosmic Watches possess advantages over Geiger counters already detailed in Section 3.3, though the latter tech has been successfully employed in Hessdalen, a hotspot known for its unexplained lights in rural Norway studied by collaborating Norwegian and Italian researchers~\cite{Hessdalen2001}. They found evidence~of above-average radioactivity possibly geomagnetic in nature~\cite{Teodorani}.

\begin{figure}[ht!]
\centering
\includegraphics[width=0.99\textwidth,clip]{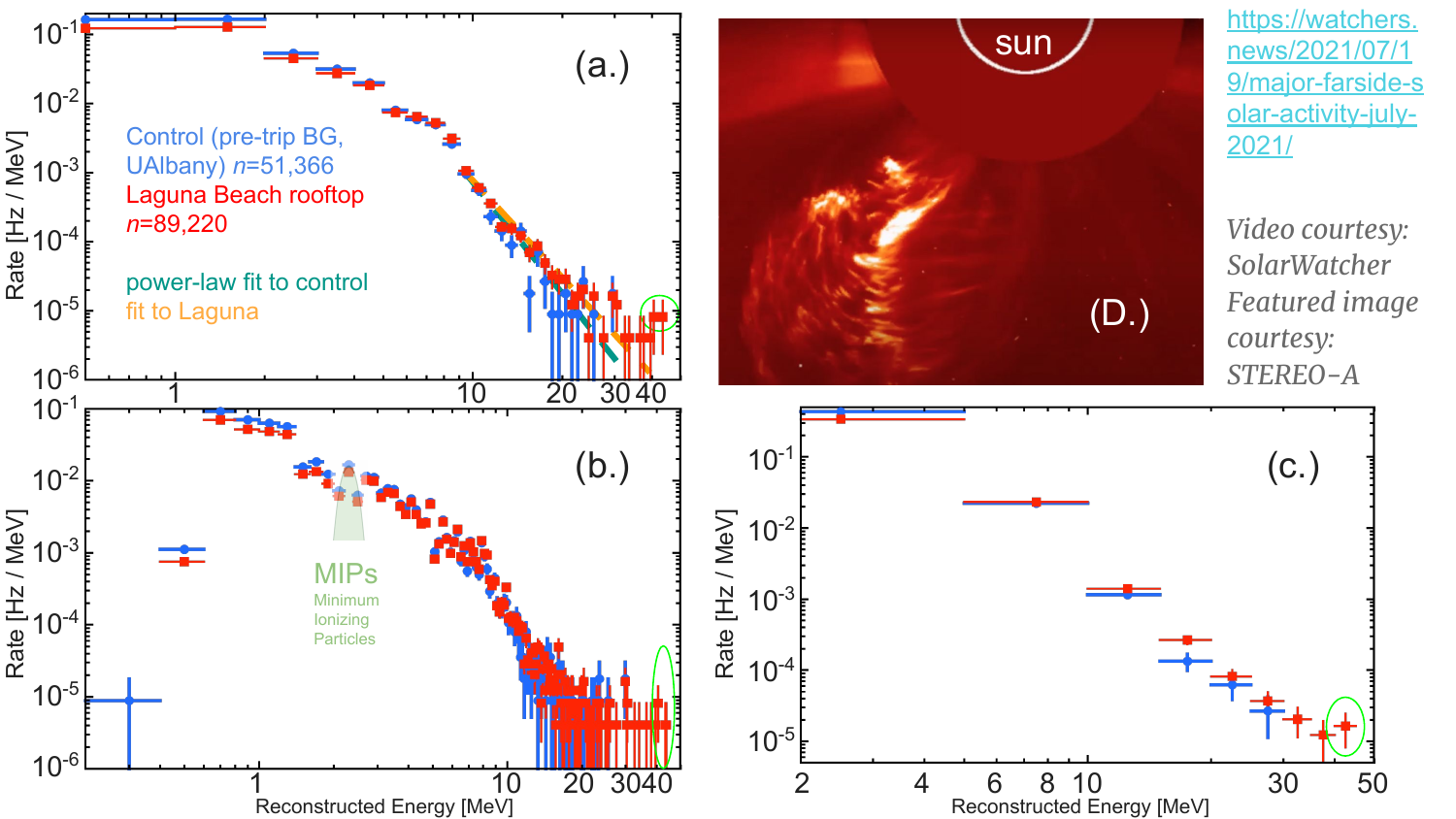}
\caption{(a.) Pre-expedition control data as cyan diamonds and Laguna Beach, California data as red squares from the MIT Cosmic Watch, with a coarse binning. Power laws, as expected, are fit to them above the cut-off for terrestrial radiation. (b.) The~same~data sets but with finer binning, enabling one to see the roll-off at low energies due to the device's efficiency, and the feature (in green) near 2 MeV which corresponds to $\sim$2~MeV/cm minimum ionizing particles (MIPs) from the natural background of cosmic rays, primarily muons, which were used to determine the energy scaling from raw SiPM scintillation pulse area or ADC counts. (A peak is easier to see in earlier data~\cite{Axani_2018,axani2019physics,Axani_2017}).~(c.)~The~same data again but with an in-between binning. Post-trip control data exist but with a replacement unit, due to the original one breaking down, so they are not shown due to being an unreliable comparison (newer versions have a different \textit{i.e.}~higher efficiency). In plots (a.-c.), a light green oval encircles the four events above 40 MeV $\alpha, \beta, \gamma, \mathrm{and}~\delta$. (D.) Photograph of solar activity at the time of the UAPx trip. It cannot explain those four events, but may explain some higher-rate events, even though this activity was on the~far~side~of~the Sun, pointed away from Earth, as a result of indirect ``backwash'' (not a~strong~explanation).}
\vspace{-0pt}
\label{Fig10}
\end{figure}

Our own high-rate events will be in a future article, on novel applications of the Cosmic Watch. For high energies ($E$s), at least four (single-interaction) particularly ambiguous events were identified, visible in Figure~\ref{Fig10} even with different binning, especially in comparison with the control data of only 40\% less livetime: two at 43 MeV (labeled as $\alpha$ \& $\beta$ in Appendix 4) and~two~at~40 MeV ($\gamma$ and $\delta$). As a result of the intrinsic features of MIT's Cosmic Watch, the particle type is not known, so these are only electron-equivalent $E$s,~due \textit{e.g.}~to quenching factors for nuclear recoils like those from cosmic-ray neutrons. They may also be in MeV/cm, not MeV, if triggers were part of long-track vertical particles passing through 1~cm of plastic scintillator.~In~spite of the potential excess near 40~MeV in Figure~\ref{Fig10} compared to control data, by themselves these events are not statistically significant (2-3$\sigma$ max $cf.$~control data or either fit), and are explicable as cosmic rays from astrophysical phenomena. They routinely exceed 40~MeV by many orders of magnitude.~But, one coincidence caught our attention, as we discuss next -- the $\alpha$ event seemed to be close in time to at least one UFODAP video (3:59:24 am, Section 5.1).

We can estimate the ``accidental coincidence'' rate or $R_{AC}$ for videos with high-energy detections at/near $\sim$43 MeV, by applying the standard formula for this from particle physics, for a pair of measuring devices:

\begin{equation}
R_{AC} = 2 \; r_{UD} \; r_{CW} \; \Delta t.
\label{eqn:3}
\end{equation}

\noindent
The detection rates $r_{UD}$ and $r_{CW}$ for the UFODAP and~Cosmic~Watch~were:

\begin{equation}
\begin{split}
    r_{UD} = (1,716~\mathrm{files})~/~(396,300~\mathrm{s}) = 0.004330~\mathrm{Hz} \\
    r_{CW} = 1.65~\mathrm{x}~10^{-6}~\mathrm{Hz}
\end{split}
\label{eqn:4}
\end{equation}

\noindent
Eq.~\ref{eqn:4} top is rate, but looking at the inverse the mean is 176s between~videos (so, 117s is sensible; 11s is an outlier, p.~17). Eq.~\ref{eqn:4} bottom comes from~the~appropriate bin in Fig.~\ref{Fig10}, accounting for~$E$ resolution. While the video closest in time to event $\alpha$ was 0.267s~long,~we conservatively set $\Delta t=60$s, our best retroactive estimate of the maximum time~difference amongst devices. This leads us to a key lesson learned, already mentioned in Section 4, of the need for second-level time synchronization for all equipment, to permit precision calculations of coincidence timing. (Subtracting the 3:59~am from Section~5.1 from the timing of event $\alpha$, 4:01 am, yields 127s, a similar value.) The result from Eq.~\ref{eqn:3} is then only 0.9~$\mu$Hz. Superficially, this appears to be a remarkably low rate; however, the total livetime for which both the UFODAP and UAPx's Cosmic Watch were operational concurrently was 194,895~s, leading to an expectation value of 0.2 accidentally-coincident events for the week.

A na\"ive calculation would say that 0.2/1716 is a $10^{-4}$ fractional chance or 0.01\% odds ($>3.5\sigma$ one-tailed or two-tailed normal probability distribution), but this would neglect the large number of Cosmic Watch events as well as the total livetime. A statement like this would have required better synchronization to reduce $\Delta t$, with the SiPM pulse widths and video~times~already short, or a less sensitive UFODAP trigger. The more correct calculation follows: the expectation value (average) for the number of accidentally-coincident events, for data collection lasting time $\tau$, is $\langle$AC$\rangle = R_{AC} \tau$, with Poisson probability

\begin{equation}
P = \frac{\langle AC \rangle ^k}{k!} e^{-\langle AC \rangle}~\approx~\langle AC \rangle~\mathrm{for}~\langle AC \rangle \ll 1~\mathrm{and}~k = 1
\label{eqn:5}
\end{equation}

\noindent
of happening $k$ times during the trip. By taking into account the duration of an excursion (or, the total number of detections) this eliminates the mistake in assuming, for example, that a ``1 in a million'' detection is anomalous when one has~1 million or more data points (the ``Look Elsewhere Effect.'') In~the full treatment, Eqn.~\ref{eqn:5}, we see the $P$ of our coincidence was $\approx$ 20 not 0.01\%.

For simplicity, if all triggering rates are sufficiently small, so they are~well modeled by Poisson functions, and each sensor in usage can be approximated as having a temporal width (relevant time scale) much shorter than the time synchronization uncertainty, we recommend utilizing Equations~\ref{eqn:3}~and~\ref{eqn:5} in~all multi-modal studies of UAP. If three or more sensors are employed~by~a team on an expedition then we recommend generalizing Equation~\ref{eqn:3} to $N$ units with unique backgrounds $r_i$ (and/or public data sets) and resolution $\Delta t$~\cite{PhysRev.53.752}

\begin{equation}
R_{AC} = N \prod_{i=1}^N r_i \Delta t^{N-1}
\label{eqn:6}
\end{equation}

As a result of Eqn.~\ref{eqn:5}, our initial ``ambiguity'' is ultimately not classifiable as an anomaly statistically speaking when applying the two-device version of Equation~\ref{eqn:6}. There are systematic effects that may decrease its improbability further, like consideration of the existence of events $\beta, \gamma$, and $\delta$ of comparable energies as $\alpha$ lacking in similar UFODAP observations, the smaller exposure time for background data, and a lack of a non-speculative/exotic mechanism which would link a correlation with causation.
That said, there are also other considerations that may increase it if different choices are made in analysis:
    \vspace{-2pt}
\begin{itemize}
    \item Counting all videos in the coincidence calculation, instead of just those within the Cosmic Watch livetime, and only those with inexplicable (at least initially) video frames, \textit{qualitatively} ambiguous in a single sensor
    \vspace{-2pt}
    \item The overall cosmic-ray rate should have decreased at the lower latitude: Laguna Beach \textit{cf.}~Albany. (This oddness is the clearest in Fig.~\ref{Fig10}c.)
    \vspace{-2pt}
    \item There is another high-energy event, from 3:57:19 am, labeled as dalet in the table in Figure~\ref{Fig19} (22.8~MeV, expected to occur at $2\times10^{-5}$~Hz rate) that may not only be correlated with another UFODAP recording, but be related to the same incident under discussion. Multiplying 0.2, the highest probability we found for $\alpha$ overlapping with the UFODAP, with the probability of 0.1 for the 22.8~MeV event to occur within 3 seconds of a different video = ($2\times(4.33\times10^{-3})\times(2\times10^{-5})\times3\times(2\times10^5)$), along with a probability of 0.02 for those two radiation measurements occurring within 4 minutes of each other ($2 \times (2\times10^{-5}) \times (1\times10^{5}) \times 240 \times (2\times10^5)$), yields $4\times10^{-4}$ ($>3\sigma$ again). But, this may be~an~unjustified multiplication of a long string of probabilities, if causal linkages cannot be established for treatment as a single incident (even if speculative).
    \vspace{-6pt}
    \item Other potential corroborations, such as unusual radiation readings at global monitoring stations, may be relevant if near-simultaneous~\cite{MADAR}.
    \vspace{-6pt}
\end{itemize}

A simulation took into account the UFODAP trigger rate, Cosmic Watch rate in relevant energy bins, lengths of UFODAP recordings, pulse widths in the Cosmic Watch, synchronization uncertainties, uncertainties in all mean quantities, and all effects above. The result was $>3\sigma$ significance once again, for an accidental coincidence of three videos with two high-energy events.
\vspace{-10pt}
\section{Discussion}
\vspace{-5pt}
In light of the possibilities, our most intriguing event appears almost by definition to be ``ambiguous''; changing interpretations change the statistical significance. That has inspired us to recommend a general plan~for~the~field. We suggest (scientific) UAP researchers adopt the following conventions: \textit{An \textbf{ambiguity} requiring further study is a coincidence between two or more detectors or data sets at the level of $3\sigma$ or more, with a declaration of genuine \textbf{anomaly} requiring (the HEP-inspired) $5\sigma$, combining Equations~\ref{eqn:6}~and~\ref{eqn:5}.} (HEP = High-Energy Physics.) Coincidence here is defined as~``simultaneity'' within the temporal resolution, and spatial when germane. This way,~one~rigorously quantifies the meaning of \underline{extraordinary} evidence, in the same way~it has been done historically by particle physicists, who have established a very high bar to clear. The statistical significance must be defined relative to a~null hypothesis, in our case accidental coincidence, combined with causally-linked hypotheses, like cosmic rays striking camera pixels.

For cases where significance is difficult to determine, we recommend defining ambiguity based on the number of background events expected, where 1 event is the borderline: \textit{e.g.},~if $<1$ event is expected to be near-simultaneous for a particular pair of sensors, but $\ge1$ events are detected, they should each be inspected, as time permits, especially qualitatively ambiguous incidents.
\vspace{-10pt}
\section{Conclusion}
\vspace{-5pt}
UAPx mounted an expedition in July of 2021 to a (suspected) UAP~hot-spot with UAlbany SUNY physicists. Exotic ideas like UAP radiation were entertained, balanced by a strenuous pursuit of skepticism. With one possible exception, ambiguous observations ended up being identifiable. At this point, none can be classified as true anomalies, although further study of remaining ambiguities may alter this conclusion. The greatest successes from this work included equipment stress-testing in the field and creation of new software of broad applicability, the only one of its kind for IR to the best of the authors' knowledge, fusing human-driven QA with interpretable AI / ML, not relying mainly on one or the other, for UAP work. Our valuable lessons apply to any future field work, conducted by us, our contemporaries, or future scientists.

We recommend at least two of each type of sensor, and 2+ distinct sensor types. The most significant new recommendations made for this growing field were those for establishing quantitative rigor in the definition of ambiguities vs.~anomalies. Our results so far are best labeled ``null,'' but science's history teaches us the value of such results~\cite{10.1088/2053-2571/ab3918ch7}, and of robust eliminative deduction. Any new excursions, to Catalina for reproducibility, and/or elsewhere (such as to Yakima~\cite{Akers2,Akers1}), must include improvements to equipment/methods,~recognizing others' past work.
\vspace{-9pt}
\section{Future Work}

Here we address questions related to scaling up an instrumented study to richer, multi-faceted environments, connecting previous separate threads~and forging overarching plans, for present and future UAP researchers to read.

Two (separated) setups (minimum) per selected site is valuable, but years are required for replicable, non-null results. Communication should be redundant (cell, internet, radios) and all equipment mirrored at sub-sites, including in output format of data, written to RAID array. The clock synchronizations must be reviewed, and updated, periodically for all devices, as well as careful measurements performed of unit positions and view angles, if applicable, and all required restarts logged. Taking these steps should ensure that data~fusion between sub-sites is possible and practical, making target distance, size, and speed all calculable via simple math. The establishment of geographic control areas -- for non-hotspot data-taking -- as first mentioned by bullet point~(3)~at the start of Section 2, is one more requirement, for positive~(target-rich)~and negative (target-poor) axes. The first goal can be met close to a bustling~airport, with computer training~\cite{Cloete}; the second, in a quiet rural area. UAPx is already discussing possibilities within practical driving distance of UAlbany in NY state and in neighboring areas. In addition, it is considering the dual-staged process suggested by Watters~\cite{Watters_2023} of evaluating hotspots with small, inexpensive cameras first prior to commitment to a longer-term stationing.

As implied by Sections 3.4(C.) and 4, more than one deployment of electronics is ideal, not immediately adjacent to one another, to reduce electromagnetic interference~\cite{MADAR,Akers2,Maccabee}; E\&M effects must decrease in strength with distance from the source or distance squared or cubed. Batteries using different chemistries, such as lithium and sodium, and non-battery power sources such as diesel generators, could help in the attempt to combat the historically claimed incapacitance effect of UAP on instruments. The greater the number of modalities one possesses, beyond physical imaging and beyond~what~UAPx and others have used, the better. In some cases, avoidance of  devices digital in nature altogether is possible, \textit{e.g.}~supplementing one's radiation detectors, like a Geiger counter and a Cosmic Watch, with analog tech like the simple electroscope, which should not be prone to the same kind of interference.

Lastly, it is important to highlight that a deductive approach for identification is likely best, a process of elimination of possibilities. Even if this does not identify a UAP, knowing what it isn't is itself crucial. In addition, at the very least identification of patterns may become possible~\cite{JPierson}. A reduction in reliance upon the human witness and having as many different environmental sensing modalities as possible will help side-step questions of data being in error, even when they are unusual. Only through rigorous and scientific data taking and data analysis can we hope to begin solving the UAP mystery.

\vspace{-9pt}
\section*{Acknowledgements}

The authors thank U.S.~Navy veterans and UAP witnesses Gary Voorhis and Kevin Day for assembling the UAPx team. We thank O.S.I.R.I.S.~owner and Air Force veteran Jeremy McGowan for discovering our most interesting ambiguity, and iHeart journalist Matt Phelan for his key work on the radar.

The authors acknowledge OMnium Media for providing the funding for~all the Laguna/Catalina field work. In addition, we'd like to thank Seattle~area technician David Mason for use of his FLIR cameras, night vision, and other equipment. We also thank David Altman and Michael Hall for serving as~the island team. We thank iHeartMedia for financial support,~and iHeart's podcaster/reporter Claude Brodesser-Akner. For manuscript reviews, we~are~indebted to two anonymous academics, and for FLIR review to two professionals. We also thank retired electrical engineers Candy and Ralph Segar for~efforts in analysis and review. We thank Caroline Cory, Lenny Vitulli,~Michael Soto, and the entire film crew for ``A Tear in the Sky.'' We also thank William Shatner for his refreshing skepticism. We thank a Warning Coordination Meteorologist in the NOAA/NWS's San Diego office, Alexander Tardy, several other (anonymous) meteorologists, and one (anonymous) Knuth group alum, for their radar analysis assistance. We thank Alex Garcia for media relations. We thank everyone involved with the History channel show \textit{The Proof is Out There} for covering UAPx, especially Miguel Sancho and Genevieve Wong.~We also thank (former UAlbany student) Jason Sokaris for informing us of \cite{NonUnifCorr}. We thank Prof.~Mike Cifone, head of SUAPS (the Society for UAP Studies), for his final suggestions for corrections prior to our submission to RPAS.
\vspace{-10pt}
\section*{Appendix 1: The ISS in NV Binoculars}

Some detections were easily explicable, such as the ISS in NV (no zoom). A common error in UAP reports is size estimation from a single observation without independent means of knowing distance (such as radar~\cite{Randall_2023}, or albedo if the color is known~\cite{zhilyaev2022unidentified,zhilyaev2022unidentified2,loeb2022down,zhilyaev2023unidentified}), or some reliable reference. Degeneracy between sizes and distances can be broken by knowing one of these two with certainty, independently, or by having $\ge$~2 observers. The degeneracy can be expressed mathematically for an object of~angular size $\theta$, stating distance $d$ as a function of physical size $s$ (or, vice versa) as shown in Equation~\ref{eqn:7}:
\vspace{-15pt}

\begin{equation}
s = d~tan (\theta)~\mathrm{or}~d = s~cot(\theta),~\mathrm{where}~\theta = \iota p~\mathrm{and}~\iota = \frac{p_{\mu m}}{f_{mm}}~206.265''
\label{eqn:7}
\end{equation}
\vspace{-20pt}
\begin{figure}[hb!]
\centering
\includegraphics[width=0.935\textwidth,clip]{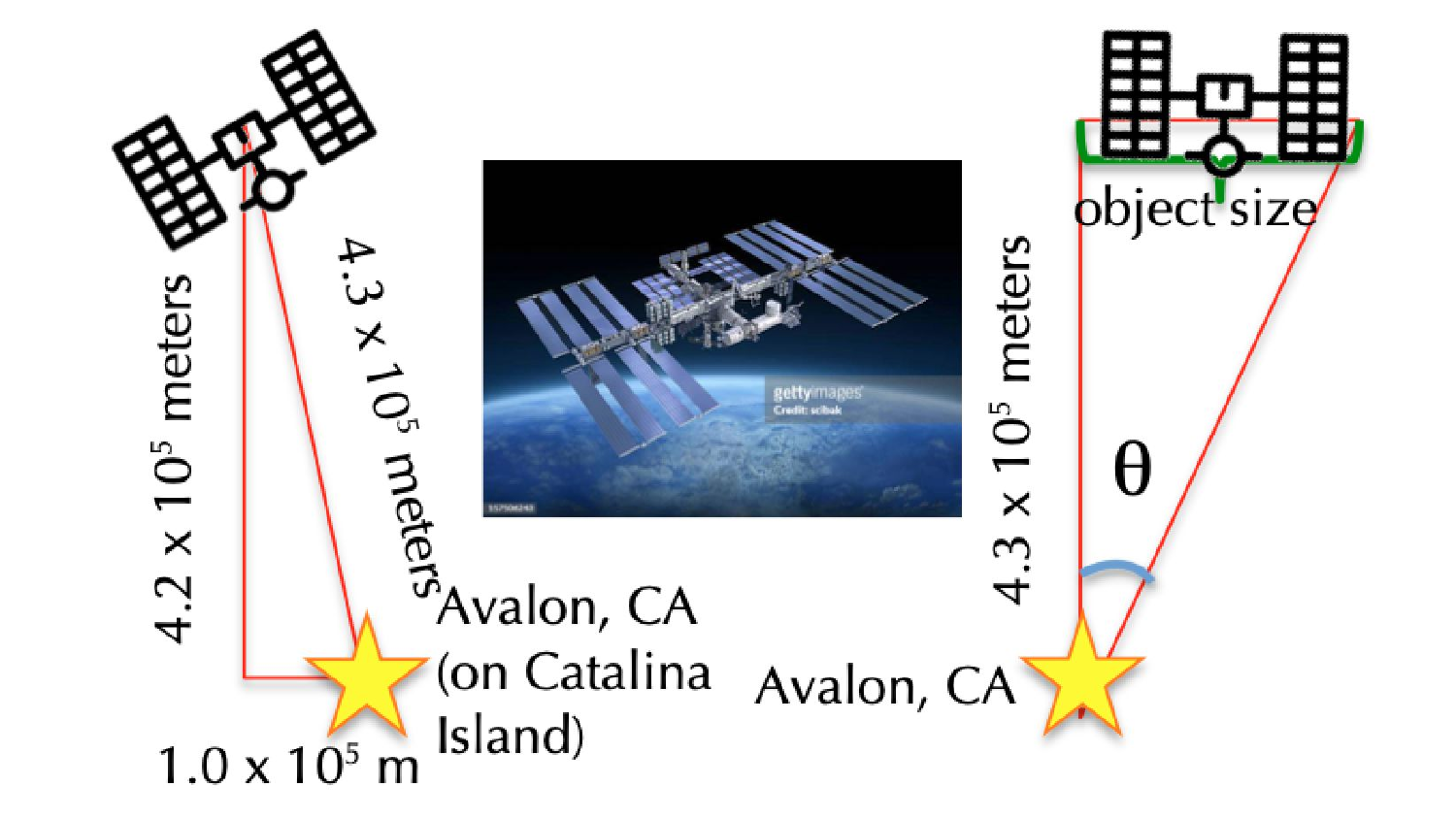}
\vspace{-15pt}
\caption{Left: A triangle possessing a base of 10$^5$~m, the distance from Figure~\ref{Fig2} between the observer position and the ISS across the Pacific Ocean. The height of this triangle is the ISS' altitude above sea level at the time. Right: A triangle formed from the hypotenuse of the left triangle, and the object in 1D, used to determine its physical size using its angular size in a camera image. As the two spotters were not a sufficient distance apart (at star) for reasonable triangulation, this is merely a consistency check, as there is a degeneracy between length and distance. The length matches the ISS within errors, if its altitude was assumed. This trigonometric exercise serves as a foundation for future UAP work.}
\label{Fig11}
\end{figure}

\noindent
Here, $\iota$ was number of arcseconds ($''$) per pixel, and $\theta$ is equal to that multiplied by $p$, the object's size in the image, in pixels, across one dimension, with $p_{\mu m}$ being the physical size of one pixel in microns within the camera's sensor, and $f_{mm}$ effective camera focal length, in mm.~(See Figure~\ref{Fig11}.) The degeneracy is clear given:~$\theta \approx s / d$. A distant UAP (at a large $d$) very great in size (large $s$) can be mistaken for something much closer (at small $d$) but also far smaller (small $s$), so that, while it shouldn't be discarded, eyewitness testimony, regarding UAP lengths and distances specifically, must be treated with caution~\cite{PowellLittle}.

Geometric calculations of the length, speed, and shape of the object spotted by night vision, further confirming its identification as the ISS, follow below. (So, they yielded another important lesson learned.) The properties~of one of the phones used to record the object (iPhone 11 pro max) were 1.4~$\mu$m pixel size and 26~mm focal length, leading to 11 arcseconds per pixel. As the object was $\sim$5 pixels across (its “core” only) this led to an angular size~determination of (11 arcseconds/pixel)(5 pixels) = 55 arcseconds = 0.015 degrees = 2.6 x 10$^{-4}$ radians.

$\theta$ = 2.6 x 10$^{-4}$ radians justifies the small angle approximation of 
tan($\theta$) $\approx \theta$ so that the object's size can be calculated as
\begin{equation}
s = (4.3~\mathrm{x}~10^5~\mathrm{m})(2.6~\mathrm{x}~10^{-4}~\mathrm{radians}) \approx 110~\mathrm{m}.
\label{eqn:8}
\end{equation}

\noindent
This result should be compared with the actual ISS size of 108 m. The size is only a match to the ISS when using the brighter central feature of the ellipse that was observed (assuming that the fuller size was generated by glow from sunlight), but changing this assumption to 10-15 pixels instead still leads to an object $O$(100~m) in size being found. 

Translating an angular to a linear velocity can be highly non-trivial, especially for a non-stationary viewing device, not deployed directly underneath a moving object in the sky. In an attempt to quantify the pixel velocity of the object in question (Figure~\ref{Fig11}), its motion with respect to a background star and the motion of that background star were both considered. Adding these components in quadrature yielded a $\sim$30 pixel/s speed. Given 110~m~/~5 = 22 m/pixel, this would imply a (tangential) speed of 660~m/s, or, ten times too small for the ISS. That being said, taking an average over the entire $2$-minute-long incident in place of a frame-by-frame analysis of instantaneous velocity is far closer. This less-crude $\omega$ estimate is based on the time it took for the object to cross the complete FOV:

\begin{equation}
\begin{split}
    \omega = (80 \pm 10^{\circ})~/~(120 \pm 10~\mathrm{seconds}) \\
    = 0.67 \pm 0.15~\mathrm{degrees/second} \\
    = (1.4 \pm 0.2)~\mathrm{x}~10^{-2}~\mathrm{radians/second}
\end{split}
\label{eqn:9}
\end{equation}
\noindent
Differentiating Equation~\ref{eqn:8} one gets
\begin{equation}
\begin{split}
\frac{ds}{dt} = (4.3~\mathrm{x}~10^5~\mathrm{m})\frac{d\theta}{dt} = (4.3~\mathrm{x}~10^5~\mathrm{m})(\omega) \\
= 6,000 \pm 1,000~\mathrm{m/s} = v
\end{split}
\label{eqn:10}
\end{equation}
\noindent
which compares favorably with the actual ISS velocity at the time, 7660~m/s.

Lastly, consider the approximate shape of the object. Our best visual~estimate of aspect ratio was about 1.5 (but definitely not 1:1). The actual~aspect ratio of the ISS is about 1.4-2 (it is orientation-dependent).

\section*{Appendix 2: FLIR ThermaCams}

The most effort was devoted to the $8 \times 75 = 600$ hours of FLIR camera recordings. Objects in thermal equilibrium with the environment are difficult to see. Any craft using conventional propulsion or animals possessing internal heating mechanisms such as avians or mammals would appear hot, but cold objects would be ambiguous.

\begin{figure}
\centering
\includegraphics[width=0.966\textwidth,clip]{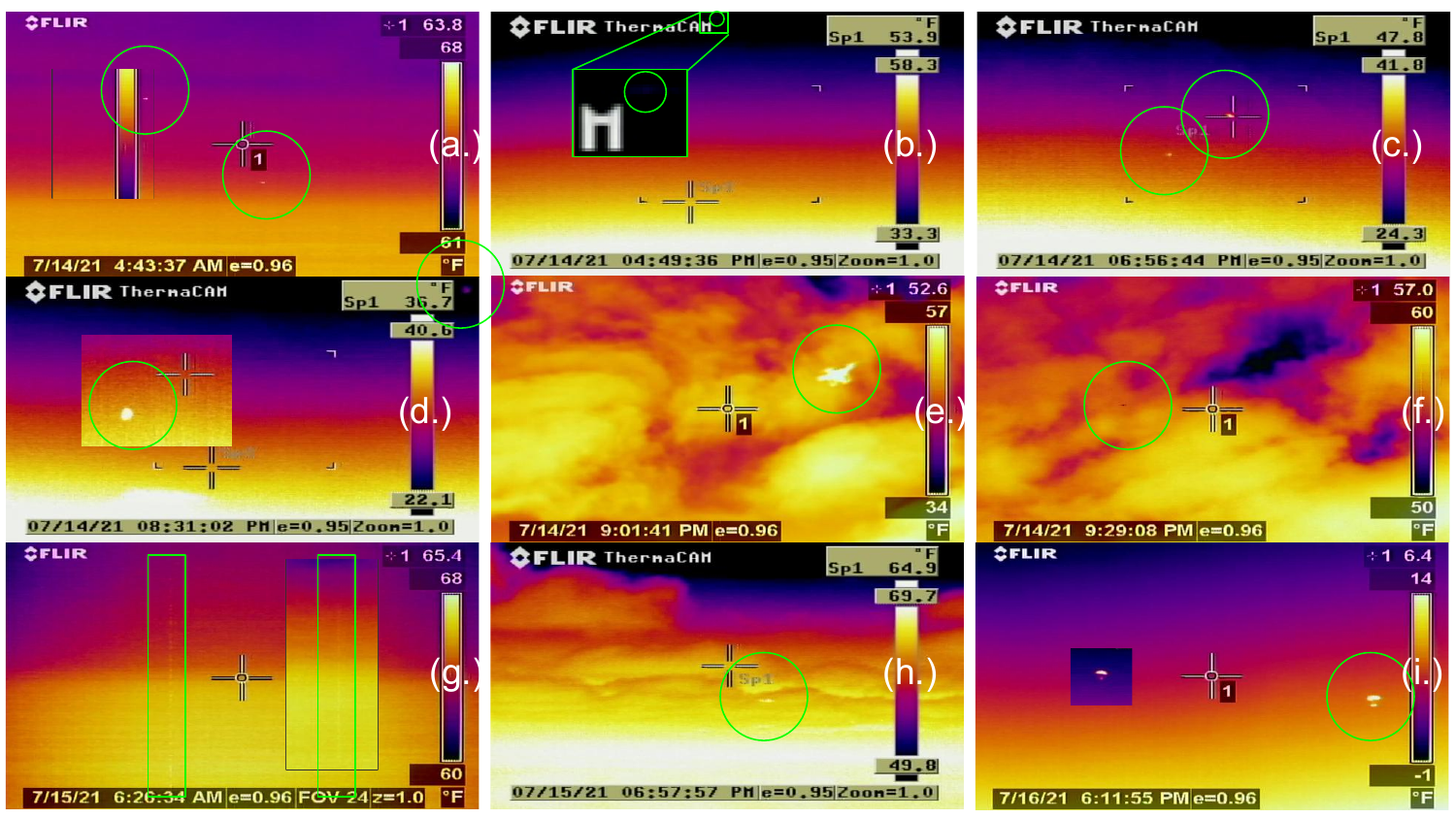}
\caption{Example FLIR events taken from different cameras, days, and times, in~chronological order from left to right and row by row downward, with C-TAP detections circled or boxed. Descriptions begin with the upper left. (a)~7/14 4:43:37am: White (color meaning off-scale hot) patch that is likely a known type of camera glitch, a bolometric discharge, with inset of an example of another one in a different position a few minutes later, also only lasting a few frames. (b)~7/14 4:49:36pm: Faint pixel cluster which traversed the~screen in a straight-line path across the top, reading as very cold, with zoom-in inset for~easier viewing. This is an example of an event not caught through human viewing, due~to~being both faint and at the screen's edge. This was most likely an international~flight. (c)~7/14 6:56:44pm: A pair of airplanes caught in the same frames, one landing and one taking~off from an LA-area airport. (d)~7/14 8:31:02pm: Example image of the Moon, at upper right, and in the inset the Moon seen again, at a different day and time and in a different camera unit. This example illustrates how an identical object (due to an immense~distance in this case of course) can appear at different temperatures, depending upon environmental factors, making an absolute determination of temperature from FLIR a challenge. (e)~7/14 9:01:41pm: Standard quad-copter drone, used to film roof scenes in the documentary film \textit{A Tear in the Sky}. (f)~7/14 9:29:08pm: An incident that remains unexplained, a red (false color) oblong shape which endured for approximately 3 frames. While similar in size and shape to a bolometric-discharge camera glitch, it reads cold, not maximally hot. (g)~7/15 6:26:34am: Vertical hot streak, one example of several around the same time frame, originally mistaken for rapidly-moving objects falling into the ocean, but ultimately identified as correlated (temporal) columnar noise. The inset is an example from 9:21am the same morning in the same camera. (h)~7/15 6:57:57pm: Helicopter flying below clouds. (i)~7/16 6:11:55pm: Para-sailor $O$(100 m) in front of the cameras, observed in multiple units. The inset shows the same recreating individual at the same time from a different point of view. Other interesting events not pictured include transient points of light most likely beacons on oil rigs in the channel bleeding into the IR, and a wingless ellipse which U-turned.}
\vspace{-0pt}
\label{Fig12}
\end{figure}

Two minutes-long incidents occurred in which dozens of objects seemingly ``rained'' into the ocean, heating it. The UFODAP failed to provide~any~corroborative detections for these events despite being operational and pointing in the same direction, except for single-pixel triggers suggestive of noise. An exotic answer is signature management, but all of these tracks were observed in only a single camera. They did not seem to be from dead bolometers, which would manifest as permanent white or black pixels (observed in other cameras), nor bolometric discharges, which would appear as localized clusters of white (hot) pixels. Two independent thermal-imaging experts from~different manufacturers (at Raytheon and Teledyne) consulted by UAPx, upon~review of our two relevant short FLIR video clips, arrived at the same~conclusions. Without consulting each other, both in their professional opinions concluded that what we had encountered was a common rastering glitch.~Despite~their banal interpretation~\cite{NonUnifCorr}, they are not willing to be identified due to the continuing taboo against the study of UAP. This observation, as well as other glitches and example events, can be seen in Figure~\ref{Fig12}.

\begin{figure}[h]
\centering
\includegraphics[width=0.99\textwidth,clip]{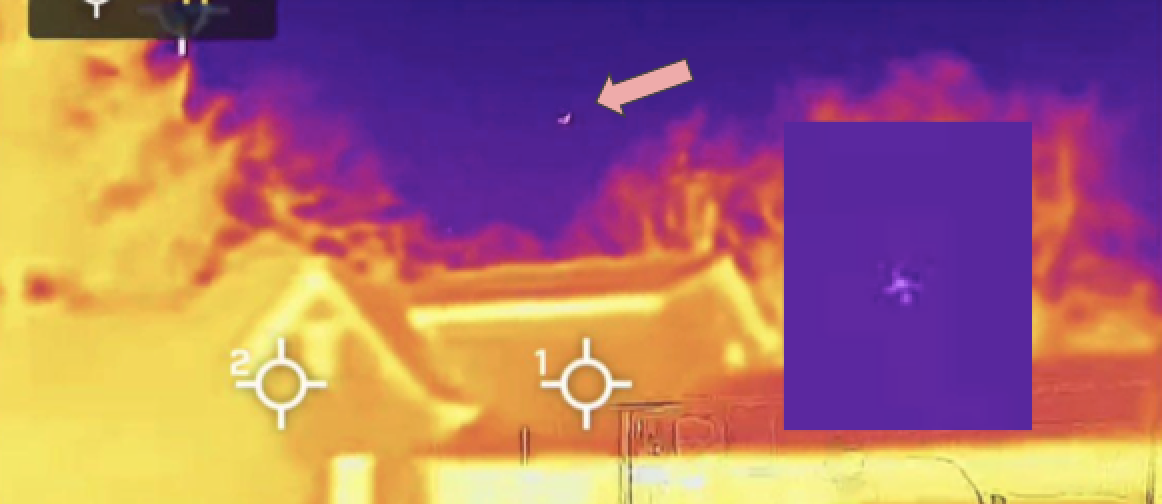}
\vspace{-5pt}
\caption{Example of the Moon (crescent phase) in the new UAPx (mini)FLIR for~smartphone. This should be compared to its appearance in Fig.~\ref{Fig12}d (left column, center row). The resolution has improved from 360p to 1080p, and the device also comes with a CCD overlay from a built-in second (visible-light) camera. Inset: Demonstration of how even distant airplanes (overhead) are more obvious in shape now, no longer generic white (hot) ellipsoids without any well-defined wings distinct from the fuselage (leading to UAP claims).}
\vspace{-15pt}
\label{Fig13}
\end{figure}

As mentioned already within Section 3.4, moving forward we will plan to use fewer, but far more portable, thermal cameras -- specifically, the Teledyne FLIR One Pro, which will be cooled down manually, and be simultaneously charging and operating to allow for multiple-hour recordings. The calibration campaign before any next expedition will involve phase transitions: of water and other substances at different distances and atmospheric conditions.~The FOV will be much more limited, but a gimbal moving to follow targets identified and tracked by the UFODAP (running different software~such as SkyHub's) will compensate for this deficiency. Figure~\ref{Fig13} has example imagery.

\section*{Appendix 3: Supplemental UFODAP Materials}

A 1/2.8-in.~2~MP STARVIS CMOS Sensor provided 20-30 fps at 1080p resolution (max) with a 125.6$^{\circ}$ (horizontal) by 73.4$^{\circ}$ (vertical) field of view.~The proprietary ``Starlight'' technology for ultra-low light sensitivity could capture details in low-light conditions down to 0.005 lux, resulting in reasonable images even for dark environments like a night sky. Depending on the focal length set, between 4.9-156.0~mm, the resolution in arcseconds was 120-3.83. The maximum possible (optical) zoom went to 32x.

This appendix has additional figures supporting those for the main video (Fig.~\ref{Fig5}, Sec.~5.1) including frames from more videos, shown chronologically.

\begin{figure}[ht!]
\centering
\includegraphics[width=0.99\textwidth,clip]{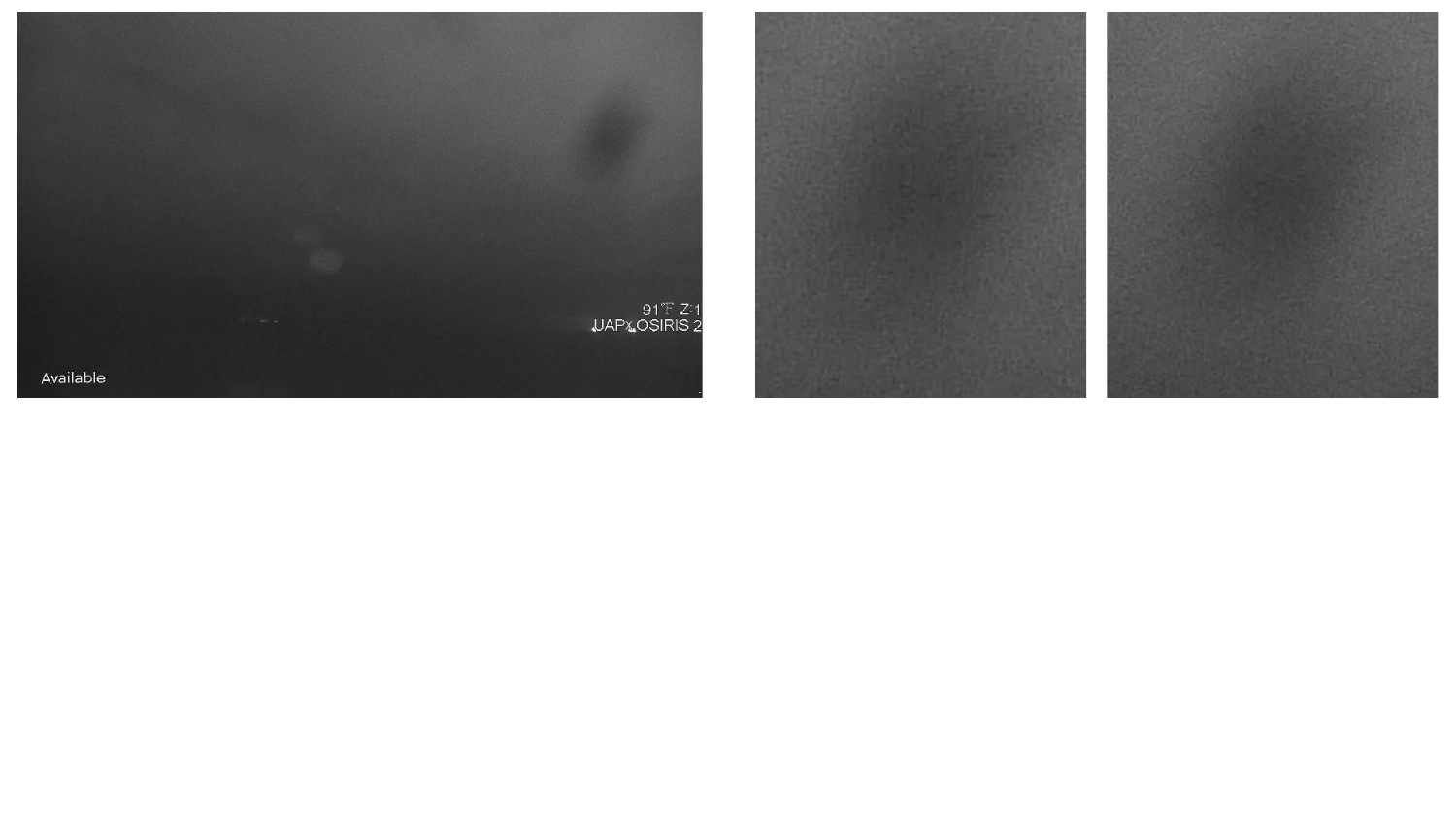}
\vspace{-120pt}
\caption{Left: The 3:57:16am spot, upper right, in the first frame of the video. Middle: A zoom in from that frame. Right: A zoom in from the final frame. No obvious changes occurred in the spot or anywhere in the field of view. (No white dots were evident yet.)}
\vspace{-14pt}
\label{Fig14}
\end{figure}

\textcolor{white}{.}
\vspace{-2pt}

\begin{figure}[hb!]
\centering
\includegraphics[width=0.99\textwidth,clip]{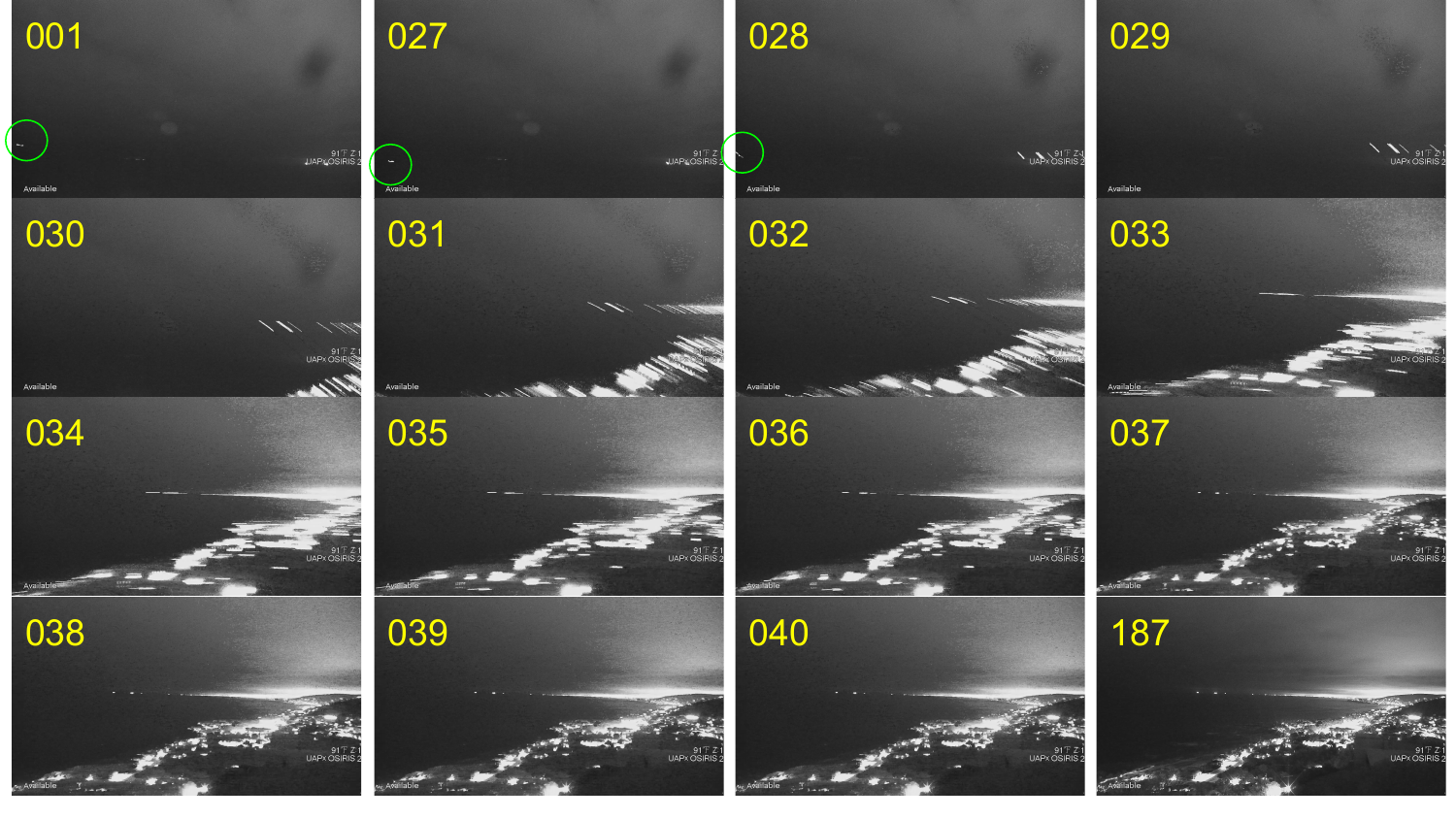}
\vspace{-13pt}
\caption{Selected images from the 187 frames of the 3:57:27am video, identified by frame number in yellow at upper left in each image. An insect unrelated to the dark spot~(in~green circles, first three images) appears to be the catalyst for the camera rotation which begins with frame 28; it is illuminated by porch lights from below. (Frames 2-26 are~omitted~due to their near-identical, nearly static nature, as well as \#s 41-186.) The dark spot from~the previous video, depicted in Figure~\ref{Fig14}, remains at upper right. During the lens’ motion,~the spot does not appear to move, at least initially. This lack of apparent motion could be due to its diffuse nature, or this could suggest it is a lens artifact, not a water droplet or a fly, especially since the spot is not at all visible in the later frames in which the sky is better illuminated. Another possibility is parallax due \textit{e.g.}~to a faraway atmospheric effect. The white dots within the dark spot first start to appear in frame 28, then appear to streak as if real, along with the city lights at lower right, although their streaks are black, not white. These dots start to merge with obvious noise (qualitatively similar to them) in frame 32.}
\vspace{-0pt}
\label{Fig15}
\end{figure}

\begin{figure}[ht!]
\centering
\includegraphics[width=0.973\textwidth,clip]{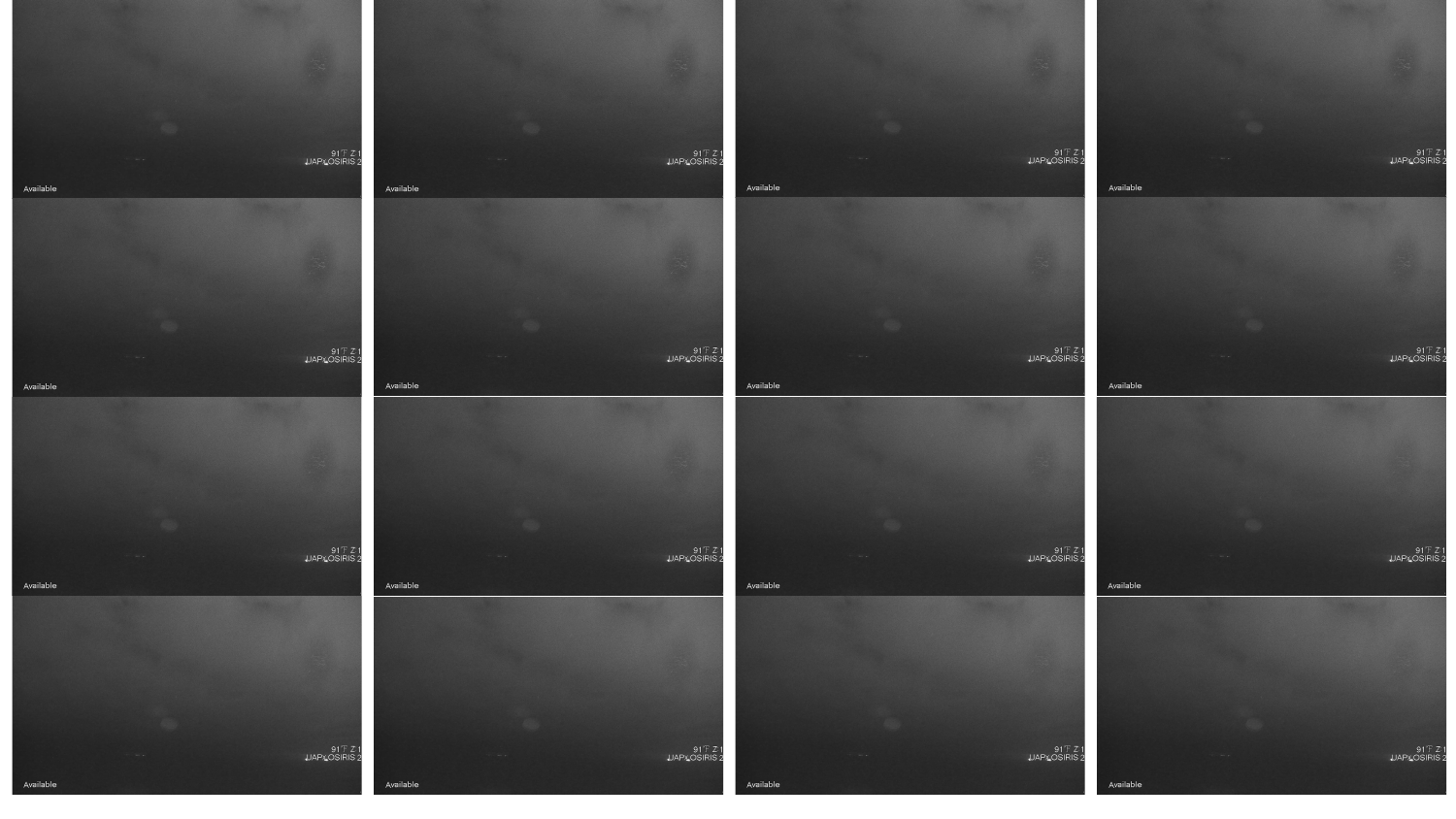}
\vspace{-15pt}
\caption{All frames of the 3:59:24 video; temporal progression is left to right~then~down, with 0.050 s in between. (This video and all others will be made available, uncompressed.) The disadvantage of an uncalibrated, triggered system is:~no images immediately preceding and succeeding these frames are available, so it is impossible to see the immediate origins of the spot and dots, or their departure. Continuous recording is the plan for the future.}
\vspace{-8pt}
\label{Fig16}
\end{figure}

An optional add-on to the UFODAP is a fisheye camera (model DH-IPC-EBW81230) with a nearly complete view of the sky. But without its~presence, the PTZ camera can self-trigger instead of being triggered via fisheye -- that's how UAPx utilized the PTZ, on a mainland rooftop. While the disadvantage is a smaller number of pixels for the same event, if the fisheye can identify a positive target through the OTDAU software it can instruct the PTZ to lock on then track that target. One or both cameras can be mounted to the top~of a vehicle for increased mobility and in our case the fisheye was permanently mounted in this fashion, to the O.S.I.R.I.S., and not coupled to the PTZ. Its purpose was a rapid deployment for providing an additional, mobile vantage point to aid in triangulation.

\begin{figure}[ht!]
\centering
\includegraphics[width=0.9\textwidth,clip]{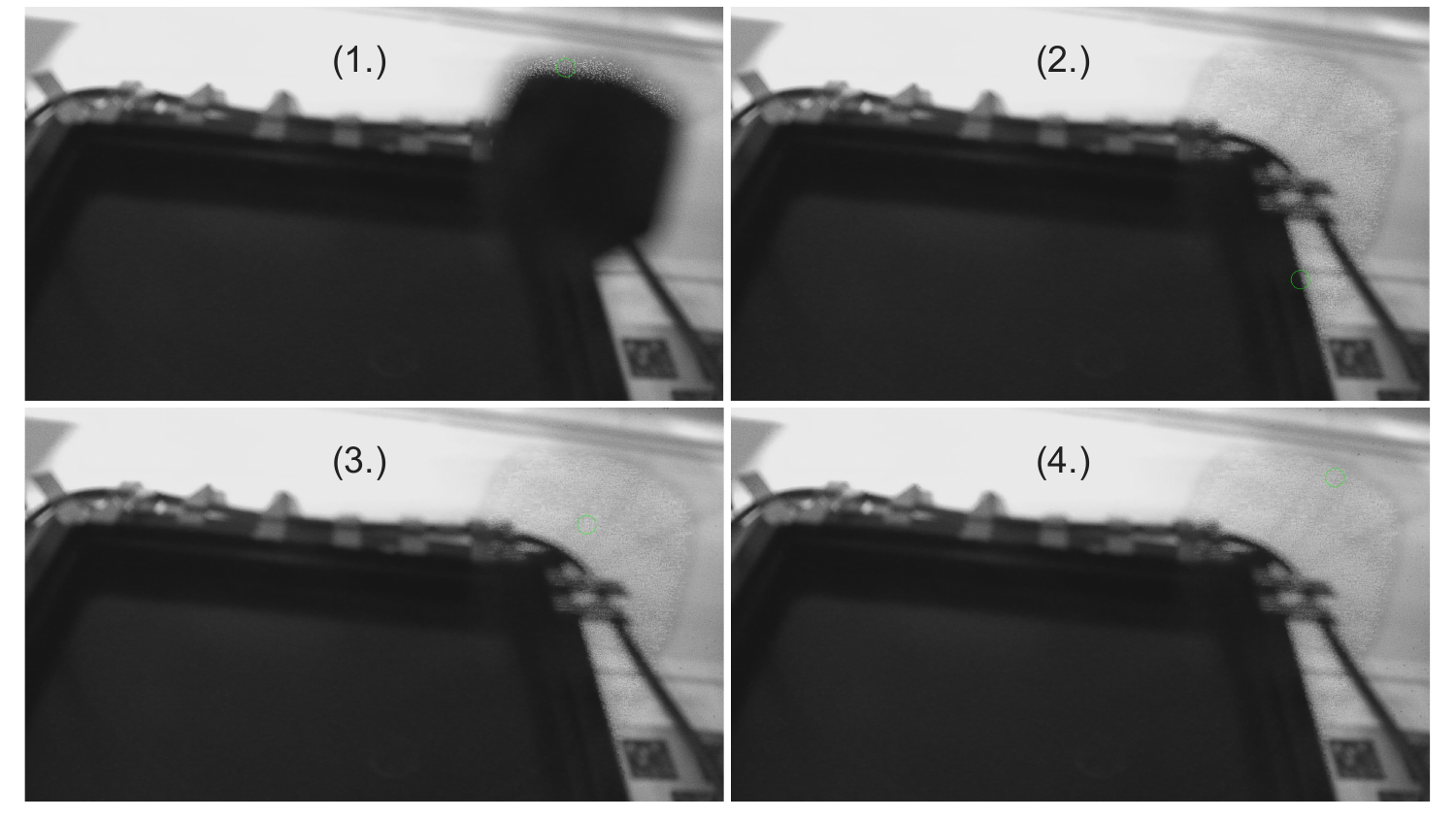}
\vspace{-12pt}
\caption{Fly stand-in (black piece of paper) out of focus on the camera dome being blown off. Dark spot fade-out (multiple-frame averaging) and white speckling were replicable~but the latter, while qualitatively similar to those from UAPx's most famous ambiguity, were either all white (8-bit value 255) or Poissonian, across multiple runs. We can't rule out~sub-Poissonian amplification of noise in the CMOS, combined with a near-field aberration like a fly takeoff and dynamic rescaling in the CMOS; however, no combination of explanations so far covers all features in order with/without slew, like single white or black dots (Fig.~\ref{Fig5})}
\vspace{-12pt}
\label{Fig17}
\end{figure}

\begin{figure}[bh!]
\centering
\includegraphics[width=0.9\textwidth,clip]{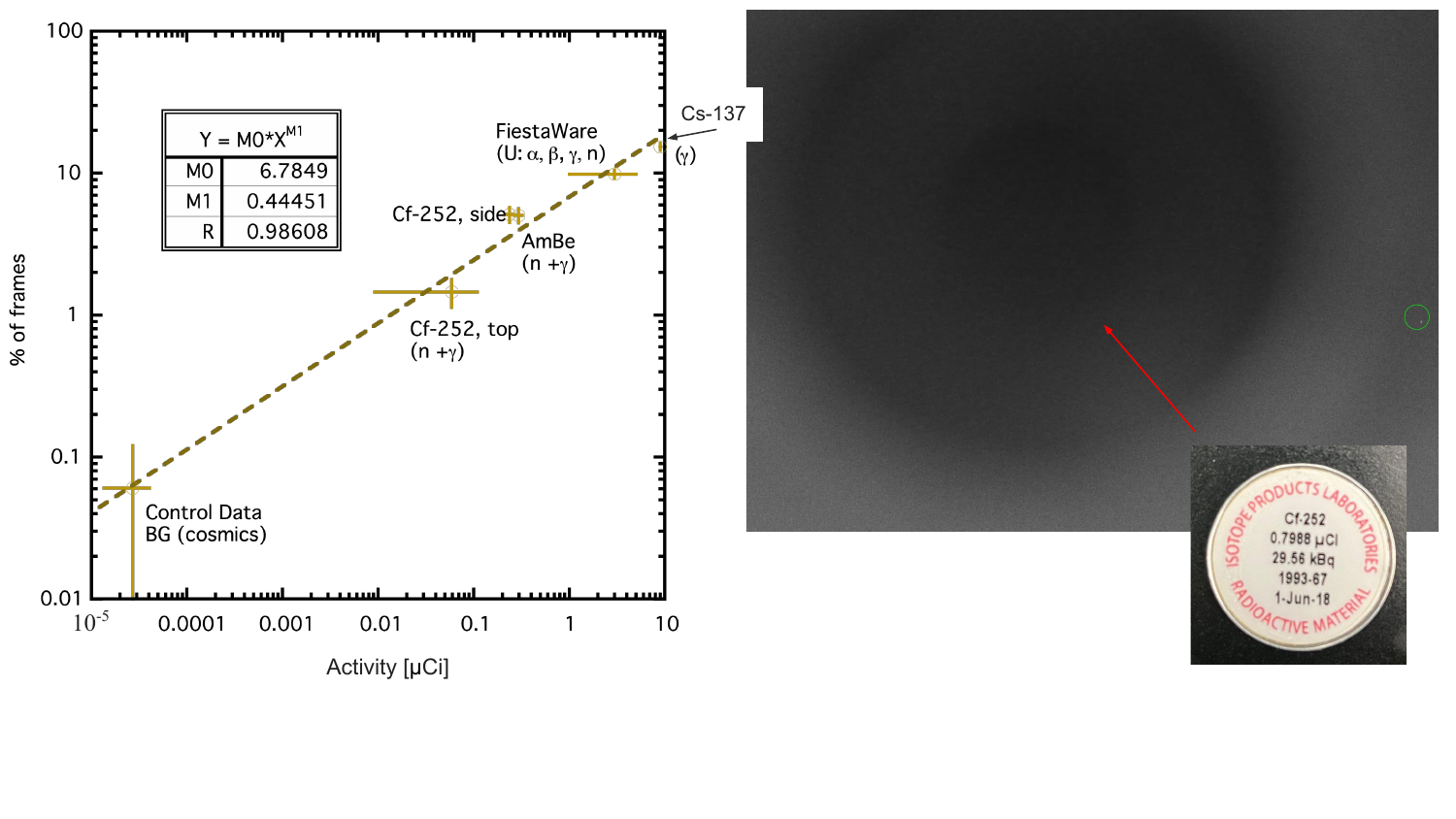}
\vspace{-44pt}
\caption{Left: The percent of frames ($>$100 implying $>$1 interactions in each frame)~with white sparkles in them, scaling with radioactivity, a novel plot for CMOS with such source variety. BG is natural background, U means Uranium, n = neutron, and the x axis is~in micro-Curies. Right: Sample frame from $^{252}$Cf calibration where it was placed on the side closest to the CMOS, with the Cf visible, and a bright dot circled by C-TAP. Inset: Source used; natural (cosmic) sources~can go much higher in effective activity, as well as energy.}
\vspace{-0pt}
\label{Fig18}
\end{figure}

\begin{figure}[ht!]
\centering
\includegraphics[width=0.95\textwidth,clip]{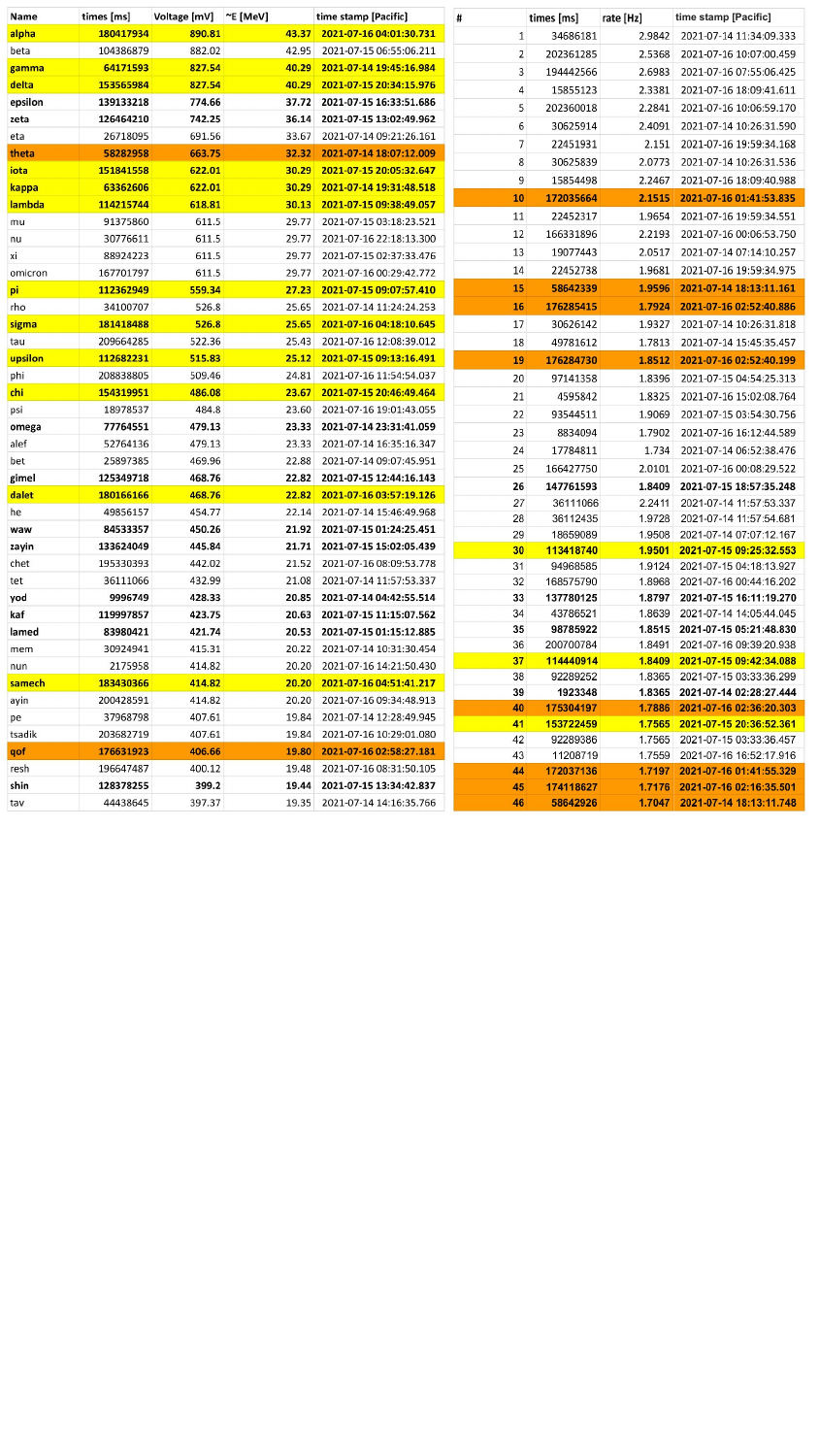}
\vspace{-295pt}
\caption{Left: List of high-energy events labeled in Greek then Hebrew letters high-low (top-bottom), converted from raw pulse area, and timestamps in ms converted into local time for the final column. Bold lines are more ambiguous, as they cannot be attributed to a CME~(Coronal Mass Ejection) or a solar flare. 2~MeV is the expected average (muons), while 8~MeV is the maximum for most natural Earthly sources of radiation. Yellow coloring indicates suspected temporal correlations of multiple detections high in energy, and orange indicates potential clusters of events high in both energy deposited and count rate. Right: A similar table but for the highest rates, numbered from highest to lowest, for the week. (These will be further probed in a later paper about Cosmic Watches for both UAP as well non-UAP research, such as solar physics.) The average was $\sim$~0.5~Hz (in non-coincidence mode) for the older version of the unit used by UAPx. The bold and non-bold fonts, and color scheme of yellow and orange highlights, have the same meaning as in the left table.}
\vspace{-0pt}
\label{Fig19}
\end{figure}

Lastly, tests were conducted at UAlbany, after the trip, with a UFODAP identical to that in CA, using only the PTZ not a fisheye, to rule in/out various mundane hypotheses for the dark oval spot, enumerated earlier. Testing was performed in a dark lab, with all lighting off and black curtains blocking daylight, to lead to the PTZ switching to low-light (gray, noisy night vision) mode, for closer comparison. Bright lights, water droplets, insect analogues, radioactive sources, and combinations of these were explored. Two example highlights of recent results can be seen in Figures \ref{Fig17} (fly) and \ref{Fig18} (radiation).

Even for hot sources placed nearby, there were only 1-2 dots per frame.

\vspace{-8pt}

\section*{Appendix 4: Table of Cosmic Watch Results}
\vspace{-10pt}

\bibliography{references.bib}

\end{document}